\documentclass[twocolumn]{aastex631}
\usepackage{float}
\usepackage{multirow}
\usepackage[para]{threeparttable}
\usepackage{enumitem}
\usepackage{hyperref}
\let\tablenum\relax
\usepackage{siunitx}
\usepackage{supertabular}

\usepackage{longtable}

\begin{document}

\title{Dust production rates in Jupiter-family comets II: Trends and population insights from ATLAS photometry of 116 JFCs} 

\author[0000-0003-4094-9408]{A. Fraser Gillan}
\affiliation{Astrophysics Research Centre, School of Mathematics and Physics, Queen’s University Belfast, Belfast, BT7 1NN, UK}

\author[0000-0003-0250-9911]{Alan Fitzsimmons}
\affiliation{Astrophysics Research Centre, School of Mathematics and Physics, Queen’s University Belfast, Belfast, BT7 1NN, UK}

\author[0000-0002-7034-148X]{Larry Denneau}
\affiliation{University of Hawaii Institute for Astronomy, Honolulu, HI 96822, USA}

\author[0000-0001-5016-3359]{Robert J. Siverd}
\affiliation{University of Hawaii Institute for Astronomy, Honolulu, HI 96822, USA}

\author[0000-0001-9535-3199]{Ken W. Smith}
\affiliation{Astrophysics Research Centre, School of Mathematics and Physics, Queen’s University Belfast, Belfast, BT7 1NN, UK}

\author[0000-0003-2858-9657]{John L. Tonry}
\affiliation{University of Hawaii Institute for Astronomy, Honolulu, HI 96822, USA}

\author[0000-0002-1229-2499]{David R. Young}
\affiliation{Astrophysics Research Centre, School of Mathematics and Physics, Queen’s University Belfast, Belfast, BT7 1NN, UK}

\begin{abstract}
Jupiter-family comets (JFCs) have orbital periods of less than 20 years and therefore undergo more frequent sublimation compared to other comet populations. The JFCs therefore represent the ideal dynamical population for investigating the dust production rates at high-cadence. We analyzed observations 
by the Asteroid Terrestrial-impact Last Alert System (ATLAS)
of 74 JFCs that reached perihelion in 2022 and 2023. The work contained in this study builds upon our previous work \citep{2024PSJGillan}, for a total of 116 JFCs over a four-year period. Using the $Af\rho$ parameter, we measured the dust production rates of each JFC as a function of heliocentric distance. We found that there remained a clear preference for JFCs to reach their maximum $A(0^\circ)f\rho$ post-perihelion, with 170P/Christensen, 254P/McNaught and P/2020 WJ5 (Lemmon) reaching a maximum $A(0^\circ)f\rho$ between 200--400 days after perihelion. However, all JFCs reached their maximum dust production within 10\% of their orbital period relative to perihelion. Fitting $A(0^\circ)f\rho$ as a function of $(R_{h})^{n}$, we measured statistically significant differences in the distribution of pre-perihelion and post-perihelion activity index $n$, with average activity indices of $-5.2 \pm 4.5$ and $-3.2 \pm 2.7$ respectively. We derived upper limits for the nuclear radii of comets 444P/WISE-PANSTARRS and 459P/Catalina as $R_{n} \leq 1.5 \pm 0.2$ km and $R_{n} \leq 1.7 \pm 0.1$ km respectively. We measured six outbursts in comets 97P/Metcalf-Brewington, 99P/Kowal 1, 118P/Shoemaker-Levy 4, 285P/LINEAR and 382P/Larson. From our four years of observing JFC outbursts in the ATLAS data, the average increase in magnitude was $- 1.3 \pm 0.8$. 
\end{abstract}

\section{Introduction} \label{sec:intro}
Jupiter-family comets (JFCs) are characterised by orbital periods of less than 20 years \citep{1997IcarLevison} and originate in the Kuiper Belt \citep{1988ApJDuncan, 1990ApJQuinn}. The orbits of these short-period comets are typically low inclination and strongly influenced by gravitational interactions with Jupiter. Due to their relatively short orbital periods, JFCs experience more frequent sublimation compared to other dynamical comet populations, making them the ideal candidates for high-cadence studies of dust production rates and activity evolution. Studying the activity evolution of JFCs is crucial for understanding how the mass-loss from this population depends on their instantaneous heliocentric distance, and how it varies over the orbit. In turn, this can have a direct bearing on the total lifetime of JFCs in the inner solar system. Repeated measurements over several orbits may allow detection of any secular evolution of activity, caused by changes in spin characteristics, nucleus surface changes, or some combination of the two.

Previous high-cadence studies of cometary activity have often focused on individual comets during outbursts, periods of heightened activity, or single perihelion passages \citep{2016IcarMiles, 2019ApJFarnham, 2022MNRASGardener, 2024arXivDobson}. Examples of high-intensity monitoring of a single comet were the campaigns on 67P/Churyumov-Gerasimenko reported by \citet{2017MNRASKnight} and \citet{2017MNRASOpitom}. While these studies provide valuable insights into the behaviour of individual comets, they lack the ability to provide comprehensive insights across a population. Conversely, investigations of large samples of objects in cometary populations such as long-period comets (LPCs), Kuiper Belt objects (KBOs), Centaurs, and JFCs have yielded population-wide trends but generally have limited information on the time evolution of specific bodies. \citep{1995IcarAHearn, 2009IcarFink, 2012IcarCochran, 2015ApJBauer, 2022PSJPinto, 2023MNRASBetzler, 2023PSJGicquel, 2023PSJDobson}. A notable example of the value of consistent survey data for identifying population-wide trends is the study of Kreutz comets by \citet{2010AJKnight}, which analyzed the light curves of 924 comets observed by SOHO.

Our work focuses on JFCs, employing consistent observational high-cadence data near perihelion to investigate the activity evolution and population-level trends. In our initial study described in \citet{2024PSJGillan} (hereafter referred to as Paper I), we measured $A(0^\circ)f\rho$ for 42 JFCs reaching perihelion in 2020 and 2021 using Asteroid Terrestrial-impact Last Alert System (ATLAS) survey data. We found that most JFCs reached their maximum $A(0^\circ)f\rho$ post-perihelion within $\sim$3 months, with two JFCs reaching maximum activity over a year after perihelion. We also calculated an activity index $n$ for the JFCs in the form $A(0^\circ)f\rho \propto R_h^n$, finding that the rate of change of dust activity pre-perihelion was greater than post-perihelion, although the small sample size meant the difference was not statistically significant.

In this work, we have used additional data from ATLAS to make high-cadence measurements of JFCs reaching their perihelion in 2022 and 2023. We have used the same methodology to ensure cross-comparisons can be readily made from our measurements. These observations were obtained to supplement and build upon the work presented in Paper I. This further analysis has more than doubled the size of our original dataset, allowing for a more robust analysis. With this enhanced dataset, we have re-examined the results from Paper I to determine whether they hold true with the larger sample, as well as searched for any new trends that may have emerged.

\section{ATLAS data and target selection} \label{sec:data}
\subsection{ATLAS} \label{section:ATLAS}
ATLAS is a survey created to evaluate the hazard posed to the Earth by asteroid impacts \citep{{2011PASPTonry}, {2018PASPTonry}} and is designed to detect small (10 -- 140 m) asteroids on their trajectory before impacting the Earth \citep{2018AJHeinze}. ATLAS is a robotic wide field survey that has been continuously operational since installation in 2016. It is currently composed of four independent units located on Mauna Loa and Haleakala in Hawaii, the South African Astronomical Observatory and El Sauce Observatory in Chile. A fifth ATLAS unit (ATLAS-Teide) is currently under testing and will be installed at Teide Observatory on Tenerife island, Spain \citep{2023snddLicandro}. 

Each of the first four ATLAS units is composed of a 0.5 m Wright-Schmidt telescope and the camera for each telescope (ACAM) provides a field of view covering 5.4\textdegree\ $\times$ 5.4\textdegree\ equating to 29 square degrees \citep{{2018PASPTonry}, {2020PASPSmith}}. The $10560\times10560$ STA1600 CCDs deliver a pixel scale of \SI{1.86}{\arcsecond}/pixel. This is larger than the typical atmospheric seeing at the sites and therefore the angular resolution is governed by the optics and accuracy of the temperature-adjusted focus, which is $\sim 2$ pixels on most nights. Using 30 second exposures, the ATLAS units can achieve a limiting magnitude of $\sim$19.7 in dark skies. The survey strategy from 2016-2024 was to take 4 exposures of each survey field with a cadence of 5-45 minutes during a night of observations. Typically, between 800 and 1000 frames were taken routinely per telescope each night. 

As ATLAS is made up of telescopes in four different locations around the globe, 24-hour night sky coverage can be achieved assuming good observing conditions, surveying $\sim$ 80 \% of the visible sky every night. ATLAS uses two non-standard broadband filters, cyan (\emph{c}) with a wavelength range of 420 nm -- 650 nm and orange (\emph{o}) that covers a wavelength range of 560 nm -- 820 nm. These filters have effective central wavelengths of $\lambda_{eff}$ = 533 nm for the \emph{c}-filter on all telescopes, and  $\lambda_{eff}$ = 678 -- 679 nm for the \emph{o}-filter depending on the telescope \citep{2018PASPTonry}. These are visible region filters that cover most of the \emph{gri SDSS} filter system. 

One of the key features and advantages of using \mbox{ATLAS} is its high cadence and wide field of view, with \mbox{ATLAS} returning to the same area of the sky much more frequently than other surveys such as the Pan-STARRS Survey for example, see \citet{{2016Chambers},{2020ApJSMagnier}}. Although \mbox{ATLAS} covers a larger area of sky than the Pan-STARRS Survey, Pan-STARRS can observe much fainter objects down to $\sim$ 22 magnitude. For solar system objects, this typically means that Pan-STARRS can observe fainter objects further out and observe comets for example over a much more appreciable fraction of their orbit, albeit at a lower cadence.

The JFCs we observed in this study exhibit motion of less than \SI{1}{\arcsecond}/pixel over the 30 second ATLAS exposures. Given this minimal displacement, the effect of motion on the photometric measurements is negligible. Photometric uncertainties in our measurements include the formal zero-point uncertainties in the each frame from the Refcat2 \citep{2018ApJTonry} calibration. The ATLAS images are calibrated via a reduction pipeline using the ATLAS \mbox{Refcat2}, which also calibrates out the slight differences in bandpasses between the telescopes to provide consistent AB magnitudes and fluxes. The survey pipeline also subtracts static sky images to create difference images, and generates a database of detected sources from those images \citep{2018PASPTonry}. An adapted version of the Pan-STARRS Moving Object Processing System (MOPS; \citealp{2013PASPDenneau}) is then used to link detections from multiple images into tracklets, subsequently identifying them as moving objects. In this study, we measured these difference images, see Paper I for further details.

\subsection{JFC target selection using expanded ATLAS dataset} \label{subsec:com_selection}
This paper describes measurements of the JFCs that came to perihelion in 2022 and 2023. This dataset expansion has more than doubled the size of our original dataset, allowing for a more robust analysis. We have a combined dataset of 4 years of JFCs from 2020 -- 2023.

For the JFCs that reached perihelion in 2020 and 2021, presented in Paper I, we were initially unaware that querying \emph{JPL Horizons} with the `time of perihelion passage' parameter did not capture all comets meeting that criterion. \emph{JPL Horizons} does not always provide the most recent orbital parameters as we anticipated. Consequently, some comets were excluded from our initial dataset presented in Paper I. However, since the choice of specific years as a selection criterion was arbitrary, primarily for the convenience of dividing the dataset into manageable subsets and the comets coming to perihelion each year are effectively randomized by their orbital periods, the exclusion of JFCs from the 2020 and 2021 datasets should not have affected the overall results. Given the size of the JFC population, it is unlikely that this omission had a substantial effect on the trends observed and discussed in Paper I.

To address this issue for the JFCs that came to perihelion in the 2022 and 2023 data, we not only selected the comets that met the criteria in \emph{JPL Horizons}, but also conducted a comprehensive review on JFC perihelia to ensure completeness. Specifically, we calculated the time of perihelion passage for all known JFCs by adding their orbital period to the most recent perihelion passage provided by \emph{JPL Horizons}. This allowed us to verify whether the perihelion occurred within the 2022 -- 2023 window, ensuring that no comets were inadvertently excluded from the dataset. This approach significantly increased our sample size from 42 JFCs in the 2020 and 2021 dataset (Paper I), to 74 JFCs in the 2022 and 2023 dataset (this work). As a result, our total sample size over the four-year period is 116 JFCs. The 74 JFCs from in this work are detailed in Table~\ref{tab:ATLAS_comets}.

For this work, we selected ATLAS images containing the predicted positions of the JFCs when predicted to have magnitude of $\leq19.5 $ by \emph{JPL Horizons}. In Paper I, we initially used a lower magnitude limit of 20.5 to account for faint JFCs and potential outbursts that could temporarily increase their brightness. However, after searching for faint JFCs near this magnitude, we found they were not detectable in the images. Based on this experience, we adopted a magnitude limit of 19.5 in this work. The resultant dataset was composed of 23,945 individual images for the 2022 and 2023 JFCs (giving  a total of 34,046 images when combined with Paper I). The median airmass from our 2022 and 2023 observations was X=1.17, with 4.2\% of images having X$>$2.0. The median FWHM of the stars in this dataset was 2.3 pixels, equivalent to \SI{4.4}{\arcsecond}.

\startlongtable
\begin{deluxetable*}{lcccccc}
\tablecaption{2022 and 2023 perihelion JFCs observed by the ATLAS survey.}
\label{tab:ATLAS_comets}
\tablehead{
\colhead{Comet} & \colhead{Number of} & \colhead{Range of observation dates} & \colhead{$e$} & \colhead{$a$} & \colhead{$q$} & \colhead{$i$} \\
 & \colhead{observations} & & & \colhead{(au)} & \colhead{(au)} & \colhead{(\textdegree)}
}
\startdata
9P/Tempel 1 &    504 &    2022 January 01--2022 December 24 &  0.51 &  3.15 &  1.54 &  10.47 \\                19P/Borrelly &    200 &           2021 June 06--2022 July 11 &  0.64 &  3.60 &  1.31 &  29.30 \\              22P/Kopff &    571 &     2021 January 08--2023 January 24 &  0.55 &  3.44 &  1.55 &   4.74 \\              44P/Reinmuth 2 &    643 &         2021 May 13--2023 January 26 &  0.43 &  3.69 &  2.11 &   5.90 \\             51P/Harrington &    107 &        2022 July 25--2022 October 31 &  0.54 &  3.70 &  1.69 &   5.43 \\        61P/Shajn-Schaldach &    368 &        2022 May 26--2022 December 27 &  0.42 &  3.69 &  2.13 &   6.00 \\ 67P/Churyumov-Gerasimenko &    596 &          2021 April 22--2022 July 08 &  0.65 &  3.46 &  1.21 &   3.87 \\              71P/Clark &    394 &   2022 February 06--2023 December 05 &  0.49 &  3.14 &  1.59 &   9.44 \\           77P/Longmore &    590 &        2022 January 26--2024 July 19 &  0.35 &  3.62 &  2.35 &  24.32 \\         80P/Peters-Hartley &     57 &          2023 April 05--2023 June 28 &  0.60 &  4.02 &  1.62 &  29.92 \\                 81P/Wild 2 &    611 &     2021 October 28--2023 October 21 &  0.54 &  3.45 &  1.60 &   3.24 \\        96P/Machholz 1 &    193 &           2022 July 20--2023 July 23 &  0.96 &  3.03 &  0.12 &  57.51 \\    97P/Metcalf-Brewington &    165 &        2021 July 19--2022 January 26 &  0.46 &  4.76 &  2.57 &  17.95 \\                 99P/Kowal 1 &    918 &     2020 December 26--2023 August 22 &  0.23 &  6.10 &  4.70 &   4.34 \\             100P/Hartley 1 &    271 &     2022 February 16--2022 August 15 &  0.41 &  3.43 &  2.02 &  25.57 \\             103P/Hartley 2 &    514 &         2023 April 05--2024 April 24 &  0.69 &  3.47 &  1.06 &  13.61 \\            104P/Kowal 2 &    369 &          2021 August 06--2022 May 29 &  0.67 &  3.21 &  1.07 &   5.70 \\     113P/Spitaler &    358 &        2021 August 18--2023 March 02 &  0.42 &  3.69 &  2.14 &   5.29 \\   117P/Helin-Roman-Alu 1 &   1291 &   2019 December 24--2024 February 02 &  0.26 &  4.10 &  3.05 &   8.70 \\       118P/Shoemaker-Levy 4 &    560 &           2022 July 20--2023 July 15 &  0.45 &  3.35 &  1.83 &  10.09 \\      119P/Parker-Hartley &    773 &          2021 June 07--2024 March 22 &  0.39 &  3.81 &  2.33 &   7.39 \\                126P/IRAS &    341 &           2023 May 06--2024 March 30 &  0.70 &  5.64 &  1.71 &  45.87 \\          127P/Holt-Olmstead &    391 &       2022 July 03--2022 December 01 &  0.36 &  3.45 &  2.21 &  14.30 \\      152P/Helin-Lawrence &    569 &      2021 March 07--2023 November 26 &  0.31 &  4.48 &  3.10 &   9.88 \\         157P/Tritton &     80 &       2022 July 28--2022 November 01 &  0.56 &  3.55 &  1.57 &  12.42 \\        158P/Kowal-LINEAR &    303 &        2021 July 05--2022 October 03 &  0.11 &  5.50 &  4.88 &   7.29 \\          170P/Christensen &    759 &          2022 July 02--2024 March 18 &  0.31 &  4.21 &  2.92 &  10.11 \\                179P/Jedicke &    511 &          2021 June 23--2023 April 21 &  0.31 &  5.93 &  4.12 &  19.90 \\                  180P/NEAT &    249 &       2022 November 20--2023 July 14 &  0.35 &  3.87 &  2.50 &  16.86 \\                  196P/Tichy &    120 &  2022 September 17--2022 November 26 &  0.43 &  3.81 &  2.18 &  19.30 \\         199P/Shoemaker 4 &    369 &      2023 March 07--2023 November 14 &  0.50 &  5.88 &  2.91 &  24.94 \\            204P/LINEAR-NEAT &    317 &          2022 August 29--2023 May 25 &  0.49 &  3.58 &  1.83 &   6.60 \\              205P/Giacobini &     71 &       2021 June 11--2021 November 04 &  0.57 &  3.54 &  1.53 &  15.31 \\                  211P/Hill &    131 &       2022 October 24--2023 April 28 &  0.34 &  3.55 &  2.33 &  18.92 \\                 230P/LINEAR &    143 &        2021 June 20--2022 January 13 &  0.55 &  3.45 &  1.57 &  15.47 \\               237P/LINEAR &    387 &     2023 January 26--2024 January 05 &  0.43 &  3.51 &  1.99 &  14.02 \\                 244P/Scotti &    529 &       2021 October 04--2024 April 09 &  0.20 &  4.90 &  3.92 &   2.26 \\                 256P/LINEAR &     11 &             2023 May 15--2023 May 24 &  0.42 &  4.64 &  2.70 &  27.62 \\                  263P/Gibbs &    144 &         2022 October 26--2023 May 27 &  0.59 &  3.04 &  1.24 &  11.52 \\      274P/Tombaugh-Tenagra &    269 &        2021 November 01--2022 May 26 &  0.44 &  4.37 &  2.45 &  15.82 \\                285P/LINEAR &    334 &     2022 August 01--2022 December 06 &  0.62 &  4.51 &  1.72 &  25.04 \\            286P/Christensen &    452 &       2022 July 21--2022 November 25 &  0.43 &  4.11 &  2.36 &  17.04 \\           287P/Christensen &    323 &        2022 June 10--2024 January 07 &  0.27 &  4.17 &  3.03 &  16.32 \\                291P/NEAT &    270 &     2022 August 17--2022 December 15 &  0.43 &  4.53 &  2.57 &   6.31 \\                 310P/Hill &    185 &   2023 September 25--2024 January 14 &  0.42 &  4.19 &  2.42 &  13.12 \\              327P/Van Ness &    539 &       2022 June 18--2022 December 30 &  0.56 &  3.56 &  1.56 &  36.25 \\                   337P/WISE &     56 &            2022 May 23--2022 July 29 &  0.50 &  3.29 &  1.65 &  15.37 \\             364P/PANSTARRS &    107 &  2023 February 14--2023 September 20 &  0.72 &  2.88 &  0.80 &  12.14 \\               382P/Larson &    330 &       2021 July 17--2022 November 29 &  0.29 &  6.19 &  4.39 &   8.44 \\                 404P/Bressi &    273 &  2022 September 28--2024 February 04 &  0.13 &  4.73 &  4.13 &   9.80 \\       408P/Novichonok-Gerke &   1067 &            2021 July 05--2024 May 14 &  0.27 &  4.74 &  3.47 &  19.36 \\            422P/Christensen &     33 &   2021 November 23--2021 December 21 &  0.51 &  6.31 &  3.10 &  39.57 \\                 431P/Scotti &     11 &   2021 September 30--2022 January 21 &  0.48 &  3.47 &  1.81 &  22.39 \\          442P/McNaught &    304 &       2022 July 10--2022 November 23 &  0.53 &  4.98 &  2.32 &   6.06 \\  443P/PANSTARRS-Christensen &    289 &        2022 March 29--2022 August 27 &  0.28 &  4.13 &  2.96 &  19.89 \\         444P/WISE-PANSTARRS &    233 &        2022 March 25--2022 August 18 &  0.57 &  3.43 &  1.47 &  22.12 \\     445P/Lemmon-PANSTARRS &    203 &        2022 July 21--2022 October 31 &  0.41 &  4.06 &  2.38 &   1.09 \\             448P/PANSTARRS &      9 &    2022 October 04--2022 November 01 &  0.42 &  3.64 &  2.11 &  12.15 \\            451P/Christensen &     17 &    2023 January 13--2023 February 21 &  0.56 &  6.34 &  2.80 &  26.48 \\            453P/WISE-Lemmon &     88 &       2022 October 23--2023 April 16 &  0.58 &  5.46 &  2.28 &  27.07 \\               459P/Catalina &    417 &         2022 August 24--2023 June 06 &  0.58 &  3.24 &  1.37 &   7.17 \\       462P/LONEOS-PANSTARRS &    126 &      2022 August 01--2022 October 03 &  0.58 &  4.89 &  2.06 &   7.03 \\              464P/PANSTARRS &    277 &     2023 February 01--2023 August 14 &  0.28 &  4.69 &  3.37 &  21.66 \\                  465P/Hill &    148 &       2023 June 28--2023 November 19 &  0.61 &  6.04 &  2.33 &  25.87 \\              466P/PANSTARRS &    126 &        2023 July 02--2023 October 19 &  0.47 &  4.04 &  2.15 &  12.24 \\         471P/LINEAR-Fazekas &    283 &       2023 July 12--2024 February 02 &  0.63 &  5.71 &  2.12 &   4.79 \\          P/2021 R5 (Rankin) &    217 &        2021 July 09--2022 January 25 &  0.31 &  4.80 &  3.32 &   7.85 \\       P/2022 C2 (PANSTARRS) &    223 &          2022 March 24--2022 July 16 &  0.44 &  6.04 &  3.37 &   9.98 \\           P/2022 L3 (ATLAS) &    395 &          2022 June 06--2023 April 10 &  0.63 &  6.51 &  2.42 &  21.54 \\        P/2022 O2 (PANSTARRS) &     42 &      2022 August 30--2022 October 14 &  0.72 &  6.31 &  1.76 &   9.42 \\           P/2023 M4 (ATLAS) &    237 &       2023 June 13--2023 December 12 &  0.28 &  5.45 &  3.93 &   7.59 \\            P/2022 P2 (ZTF) &    142 &     2022 August 26--2023 February 11 &  0.56 &  4.49 &  1.98 &  12.44 \\       P/2023 B1 (PANSTARRS) &    147 &       2023 January 13--2023 April 22 &  0.13 &  7.06 &  6.14 &  14.59 \\       P/2023 M2 (PANSTARRS) &    157 &       2023 July 13--2023 November 26 &  0.37 &  5.56 &  3.51 &  19.73 \\
\enddata
\end{deluxetable*}

\section{COMETARY ACTIVITY AND MEASUREMENTS} \label{sec:method}

\subsection[Afp]{$Af\rho$ and the dust production rate $Q_d$}\label{subsec:afrho_and_Q_d}
To quantify the relative dust activity from the observed comets, the $Af\rho$ parameter was used \citep{1984AJAHearn}. It is the product of the grain albedo, $A$, the filling factor $f$, and the radius of the photometric aperture on the sky at the comet $\rho$. The $Af\rho$ parameter is particularly useful because it only relies on the measured flux from the comet, making it a readily obtainable measure of the dust production rates. $Af\rho$ is therefore often used to compare the relative dust production rates between different comets.

$Af\rho$ is given by
\begin{equation} \label{eq:AfpFlux}
Af\rho = \frac{4 \, R_h^2 \, \Delta^2}{\rho}\frac{F_{com}}{F_\odot}
\end{equation}
where $R_h$ is the heliocentric distance in au, and $\Delta$ is the geocentric distance measured in cm. $F_{com}$ is the measured cometary flux and $F_\odot$ is the solar flux at 1 au. 
 Alternatively, $Af\rho$ can be measured in terms of magnitudes, shown in Equation~\ref{eq:AfpMag}.
\begin{equation} \label{eq:AfpMag}
Af\rho = \frac{4 \, R_{h}^{2} \, \Delta^{2}}{\rho} \cdot 10^{0.4(m_{\odot} - m)}
\end{equation}
where $m_{\odot}$ and $m$ are the apparent
magnitudes of the Sun and comet respectively in the same filter. 

The flux and corresponding AB magnitude of the Sun was derived by convolving a standard solar spectrum with the system throughput \mbox{(atmosphere+telescope+filters)} as shown in \cite{2018PASPTonry}. The apparent AB solar magnitudes in the ATLAS system are $m_\odot (c) = -26.696$ and $m_\odot (o) = -26.982$.

We used a $\rho$ = 10,000 km aperture radius for all of our measurements, as per Paper I and other studies \citep{2013A&ASnodgrass, 2016MNRASBoehnhardt, 2022IcarBorysenko, 2023PASJLin}.
Most of our measurements were made at heliocentric distances $<$ 4 au, giving an aperture radius of $>$ 3.5\arcsec. With the median angular resolution in our data of 4.3\arcsec, approximating the seeing profile by a Gaussian implies that using a 10,000 km aperture radius would result in $>$ 94\% of the flux from a stellar source being measured. 99P/Kowal 1 was the most distantly observed JFC in our dataset at 4.7 $\leq$ $R_h$ $\leq$ 5.1 au, and therefore our $A(0^\circ)f\rho$ measurements for this JFC are possibly lower limits, due to aperture losses. 

The amount of scattering depends on the size and composition of the dust particles. We measured an $A(\theta)f\rho$ value where $\theta$ is the phase angle of the comet at a given date. We corrected to a phase angle of 0\textdegree \, using the Schleicher-Marcus dust phase function \citet{2010Schleicher},  giving us a quantity of $A(0^\circ)f\rho$ which can theoretically be used to estimate the dust production rate $Q_d$, serving as an indicator of the mass-loss rate from particulate matter in kg/s. 

By assuming the dust grain velocity, albedo, density and size distribution, it is possible use $A(0^\circ)f\rho$  to calculate the dust production rate in $Q_d$ kg/s.  As shown by \cite{2012IcarFink} and \citet{2018IcarIvanova} it is a very speculative and ambiguous procedure to derive $Q_d$ from $Af\rho$. It highly dependent on the assumed albedo and size distribution, the ejection velocity and maximum size ejected will vary as a function of heliocentric distance.
Therefore, although we stress that any calculation of $Q_d$ carries significant uncertainty, we have calculated an example mass-loss rate to provide an approximate order of magnitude estimate. The relationship between $Af\rho$ and $Q_d$ is given by Equation~\ref{eq:Qd}. 
\begin{equation}\label{eq:Qd}
Q_d = Af\rho \, \frac{4\, \pi \, r_d \, \sigma_d \, v_d}{3\, p_o}
\end{equation}

Here, $r_d$ is the size of the dust grains, $\sigma_d$ is the grain density, $v_d$ is the velocity of the dust grains, and ${p_o}$ is the geometric albedo in the ATLAS \emph{o}-band. To simplify, we assumed that all of the dust grains have the same size, ejection velocity and density. We used the values from Paper I, where the geometric albedo ${p_o}=0.04$ and the grain density is $\sigma_d$ $\approx$ 1 g/cm$^{-3}$ \citep{1990ApJJewitt, 2011IcarIvanova, 2014ApJMoreno}. To place upper limits on mass loss, we also assume that we are dealing with fine, fast moving dust grains of size 1 $\mu$m grains and $v_d=300$ m s$^{-1}$ \citep{2005NaturKuppers, 2010A&AFulle}.

\begin{figure}[h!]
    \centering
    \epsscale{1.2}
	\plotone{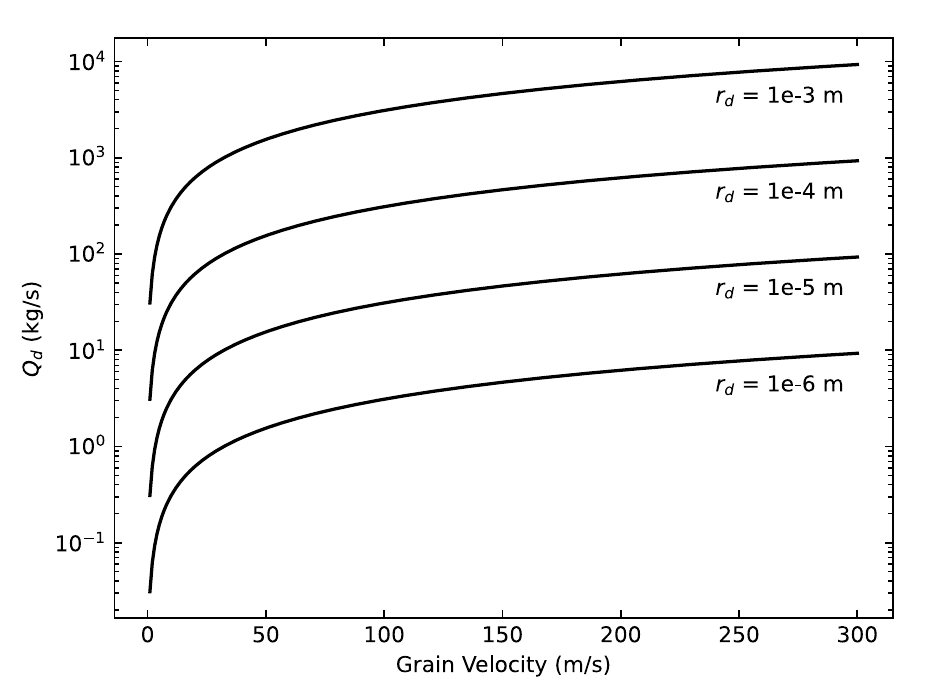}
	\caption{The variation of Qd as a function of grain velocity and particle size is shown for particles ranging from 1$\mu$m to to 1mm and dust ejection velocities between $v_d=0-300$ m s$^{-1}$ for 327P/Van Ness.}
	\label{fig:Qd_variation}
\end{figure}

The optimal approach to determine the mass loss rate, $Q_d$, for a given JFC is to consider a range of ejection velocities and a particle size distribution rather than a fixed value. In Figure~\ref{fig:Qd_variation}, we have shown how $Q_d $ varies with the dust ejection velocity and particle size for particles ranging from 1$\mu$m to 1mm and ejection velocities from $v_d=0-300$ m s$^{-1}$ for 327P/Van Ness as an example JFC (see Section~\ref{subsec:327P}). Figure~\ref{fig:Qd_variation} clearly demonstrates that the mass loss rate is highly sensitive to assumptions about the dust grain properties. This sensitivity makes it challenging to determine the mass loss rate accurately. For this reason, we have used fine, 1-micron radius dust grains with high ejection velocities only to provide an order-of-magnitude estimate for fast-moving fine dust grains as previously stated.

\subsection{327P/Van Ness as a representative JFC} \label{subsec:327P}
\begin{figure}[h!]
    \centering
    \epsscale{1.25}
    \plotone{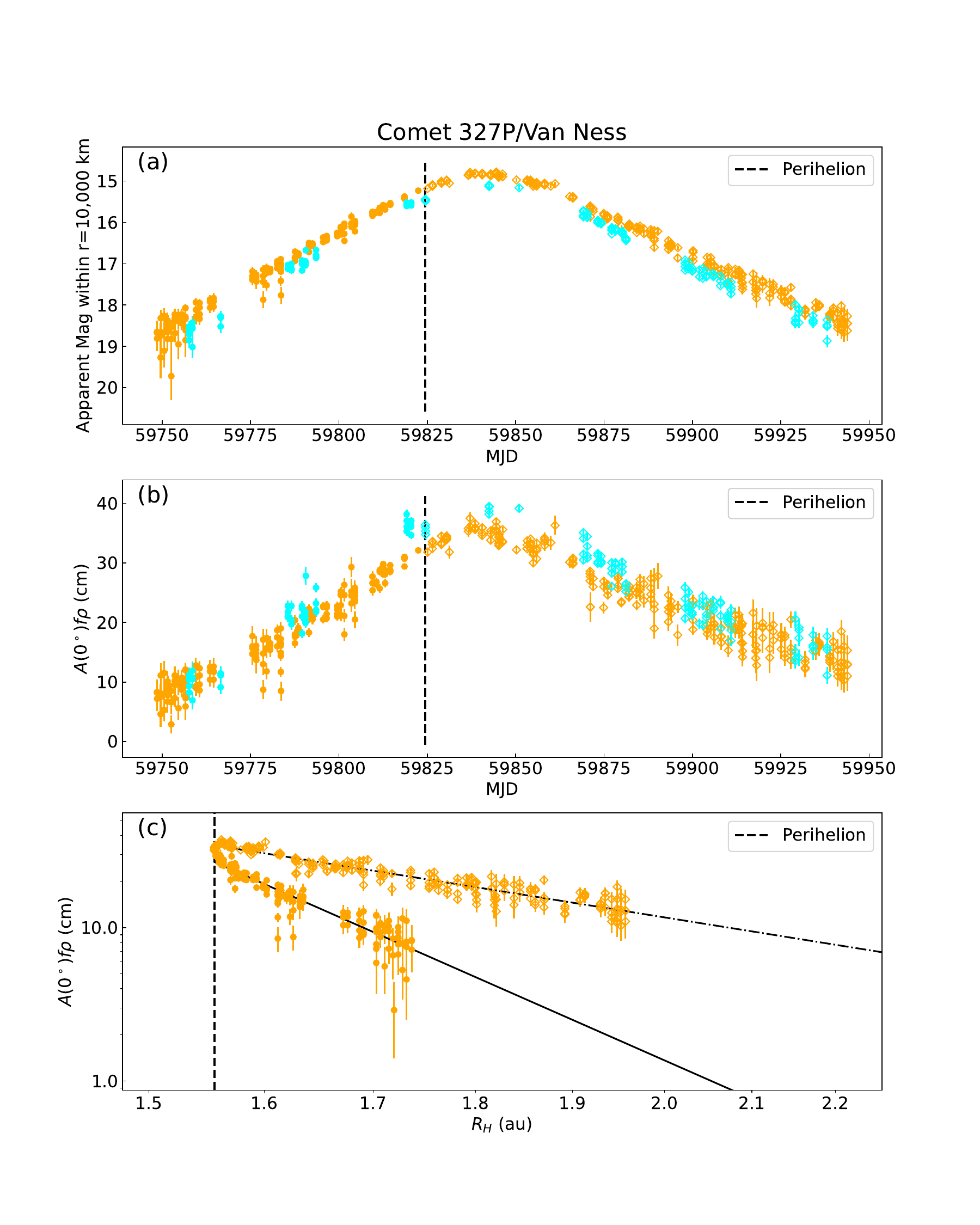}
    \caption{The 327P/Van Ness light curve is displayed in the top panel (a). The dust production parameter $A(0^\circ)f\rho$ vs time in MJD is displayed in (b). Subplot (c) shows $A(0^\circ)f\rho$ vs heliocentric distance with the solid line representing the rate of change of activity pre-perihelion and the dot-dashed line showing the rate of change of activity post-perihelion. Orange and cyan points on the plots represent the \emph{o-}filter and \emph{c-}filter respectively. Filled circular points represent pre-perihelion measurements and open diamonds represent post-perihelion. Perihelion is marked with the dashed line in each panel. The complete Figure Set is available in the online journal.}
    \label{fig:327P}
\end{figure}

Each comet in our data set was measured and analyzed using a consistent methodology. In  Figure \ref{fig:327P} 327P/Van Ness is shown as a representative example. 
This JFC was visible in 539 images taken at high-cadence through perihelion over a range of 196 days. The comet was detectable for $\sim$ 0.2 au approaching perihelion, and up to $\sim$ 0.4 au post-perihelion.

We used 1000 $\times$ 1000 pixel sub-arrays centred on each comet, positioned using the object ephemerides from \emph{JPL Horizons}. To measure the flux of each comet, we performed aperture photometry. Because the exact central position of the JFC in each image was not initially known, we centroided to precisely locate this within the frame. Using the same methods as Paper I, we minimized coma contamination in the sky measurements for aperture photometry. We used outer sky radii of $5 \times 10^{4}$ km for JFCs fainter than 15th magnitude, $1 \times 10^{5}$ km for those between 14th and 15th magnitude, and $1.5 \times 10^{5}$ km for brighter comets.

Figure~\ref{fig:327P} (a) shows the evolution of the apparent magnitude of 327P/Van Ness with time, measured in MJD. For 327P/Van Ness, the apparent magnitude within the aperture peaked post-perihelion at $\sim$10 days and at $\sim$15 mag. This JFC increased in brightness toward perihelion and decreased as it moved away as expected.

For all JFC images, the magnitudes were converted to $A(0^\circ)f\rho$ measurements via Equation~\ref{eq:AfpMag}. Outlying measurements led to manual inspection of the corresponding images. Measurement rejections occurred for several reasons, including cloud cover, the comet’s close proximity to a bright object, such as a star or satellite in the image, resulting in flux contamination within the aperture. Additional reasons for image rejection included issues with the wallpaper subtraction, the JFC being near the edge of the frame with poor image quality, or low comet brightness relative to the surrounding sky. The figures for every JFC in this study can be found in the Figure Set associated online with Figure~\ref{fig:327P}.

\begin{figure}[h!]
    \centering
    \epsscale{1.25}
	\plotone{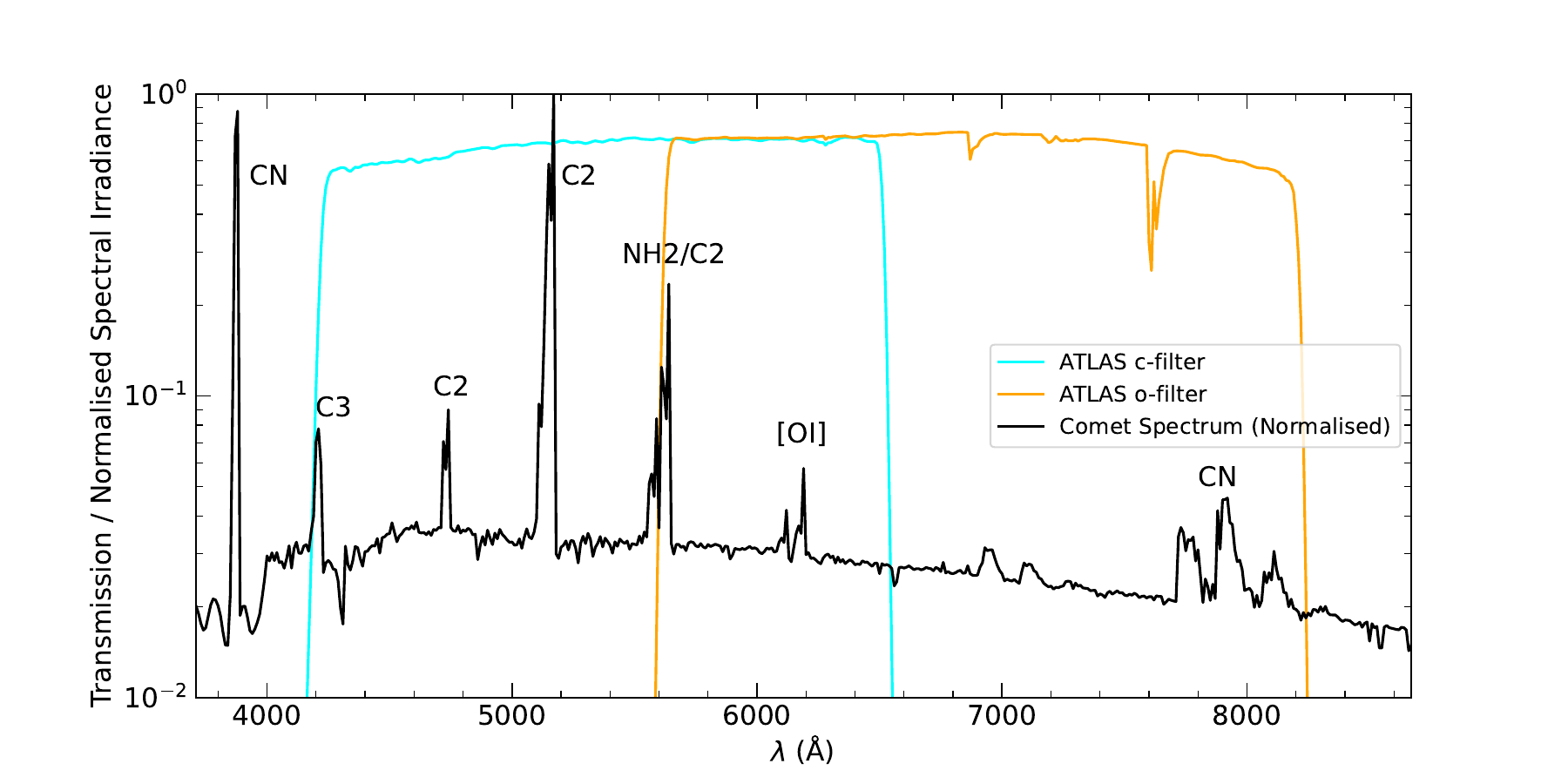}
	\caption{The transmission curves for the ATLAS \emph{c} and \emph{o-}filters are shown in cyan and orange respectively. A template comet spectrum with template gas abundances is plotted in black with the molecular emissions labelled above each feature in the spectrum.}
	\label{fig:ATLAS_and_spectra}
\end{figure}

Once any outlying measurements were rejected, each individual JFC could be analyzed. Figure~\ref{fig:ATLAS_and_spectra} shows the ATLAS transmission curves with a template comet spectrum at optical wavelengths using template gas abundances generated by NASA's Planetary Spectrum Generator \citep{2018JQSRTVillanueva}\footnote{\url{https://psg.gsfc.nasa.gov}}. In active JFCs, the \emph{c}-band could contain prominent cometary gas emission due to the C$_2 (0-0)$ band and others. The \emph{o}-band should be far less contaminated by gas emission, containing only weaker emissions such as part of the C$_2 (0-1)$ band, NH$_2$ bands, [OI] and the CN(1-0) band.

To identify comets with potentially significant gas contamination, we calculated the average $(g'-r')$ and $(g'-i')$ color from \citet{2012IcarSolontoi}. For comets not experiencing an outburst, the relative contrast between gas emission bands and the underlying dust continuum falls as heliocentric distance increases. For the subsamples of comets in this study observed at either $R_{h}>$2 au and $R_{h}>$3 au, the average colours were $(g'-r')=0.57$ and $(g'-i')=0.78$, implying no change with heliocentric distance and hence a dust-dominated spectrum. We transformed to the Pan-STARRS system using the relationships in \cite{2016ApJFinkbeiner}, and then into the ATLAS system using the expressions in \cite{2018PASPTonry}. From these data  an average dust-dominated comet coma should have a \emph{c}-band flux 6\% fainter than in \emph{o}-band. We therefore scaled all of the \emph{c}-band $A(0^\circ)f\rho$ measurements by 6\%. With this assumption of average comet dust color, the scaled \emph{c}-band $A(0^\circ)f\rho$ measurements should follow the same trend as the \emph{o}-band values, if no significant gas emission is present. If the scaled \emph{c}-band values are higher than the \emph{o}-band measurements, this implies there is excess emission in this filter that is probably due to gas, and we must be careful about interpreting the \emph{o}-band $A(0^\circ)f\rho$ measurements. An example of probable significant gas emission is shown in Figure~\ref{fig:327P} (b), where the $A(0^\circ)f\rho$ measurements in the \emph{c}-band are larger than the \emph{o}-band.

We found that there was probable gas emission in the \emph{c}-band in comets 9P/Tempel 1, 103P/Hartley 2, 104P/Kowal 2, 118P/Shoemaker-Levy 4, 327P/Van Ness and 364P/PANSTARRS. Our \emph{o}-band $A(0^\circ)f\rho$ measurements therefore may be upper limits for these JFCs and should be treated with caution. Bluer than average colors may also be due to bluer than average dust scattering. Although  we do not have the spectra to definitively determine this, we note that this blueing of the comet data generally occurs as the comet approaches perihelion, following the trend expected for gas contamination of the \emph{c}-filter. This effect can be clearly seen in Figure~\ref{fig:327P} (b),  which shows the relative dust production rate $A(0^\circ)f\rho$ versus date for 327P. Interestingly, the transformed \emph{c}-band measurements for 126P/IRAS lie below the \emph{o}-band measurements, implying that the dust in this JFC may be redder than average. 

Finally, for each comet we also measured the $A(0^\circ)f\rho$  activity index $n$ both pre- and post-perihelion where possible. Figure~\ref{fig:327P} (c) shows the \emph{o-}band $A(0^\circ)f\rho$ for 327P as a function of heliocentric distance. This is discussed in detail in section~\ref{subsec:AI} for all JFCs in our dataset.

\subsection[Detectability]{Detectability $R_h$ range for each comet} \label{subsec:detect_range}
\begin{figure*}[h!]
    \plotone{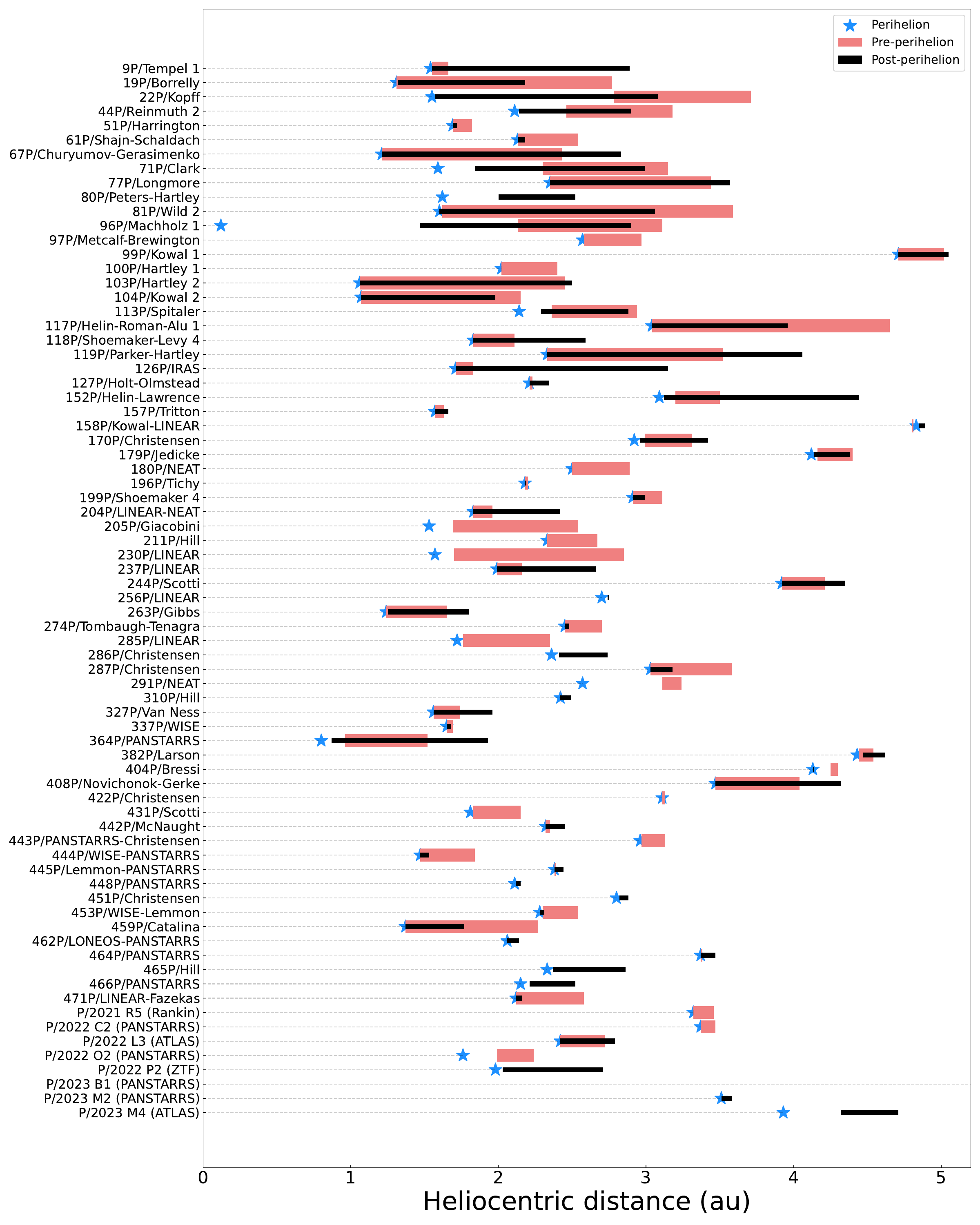}
	\caption{JFCs reaching perihelion in 2022 and 2023 the were visible to ATLAS. The blue stars represent the perihelion of each comet, pink represents the pre-perihelion $R_h$ observed range by ATLAS and black represents the post-perihelion $R_h$ observed range by ATLAS.}
	\label{fig:Rh_ranges}
\end{figure*}

Figure~\ref{fig:Rh_ranges} shows the range of heliocentric distances where ATLAS detected JFCs that reached perihelion in 2022 and 2023. This figure illustrates that the lower-numbered comets are typically detected over a broader $R_{h}$ range compared to the higher-numbered JFCs and those with provisional designations. This disparity mostly arises because the lower-numbered comets are generally brighter and thus more easily detectable. Several JFCs are only observed either pre- or post-perihelion, primarily due to solar conjunction, which temporarily obscures them. Additionally, the detectability of these comets fluctuates with varying activity levels throughout their orbits, which causes comets to fall above or below the ATLAS detection threshold of $\sim $19.5 magnitude. Note that comet 158P/Kowal-LINEAR has observations closer than its catalogued perihelion distance in Figure~\ref{fig:Rh_ranges}. This was due to a significant orbital perturbation of this low-eccentricity comet by Jupiter that was occurring during our observations.

\section{JFC POPULATION RESULTS FROM FROM YEARS OF ATLAS DATA} \label{sec:population_results}
\subsection{Maximum \texorpdfstring{$Af\rho$}{Af rho} from four years of perihelion JFCs} \label{subsec:afrho_vs_days}
\begin{figure}[h!]
    \plotone{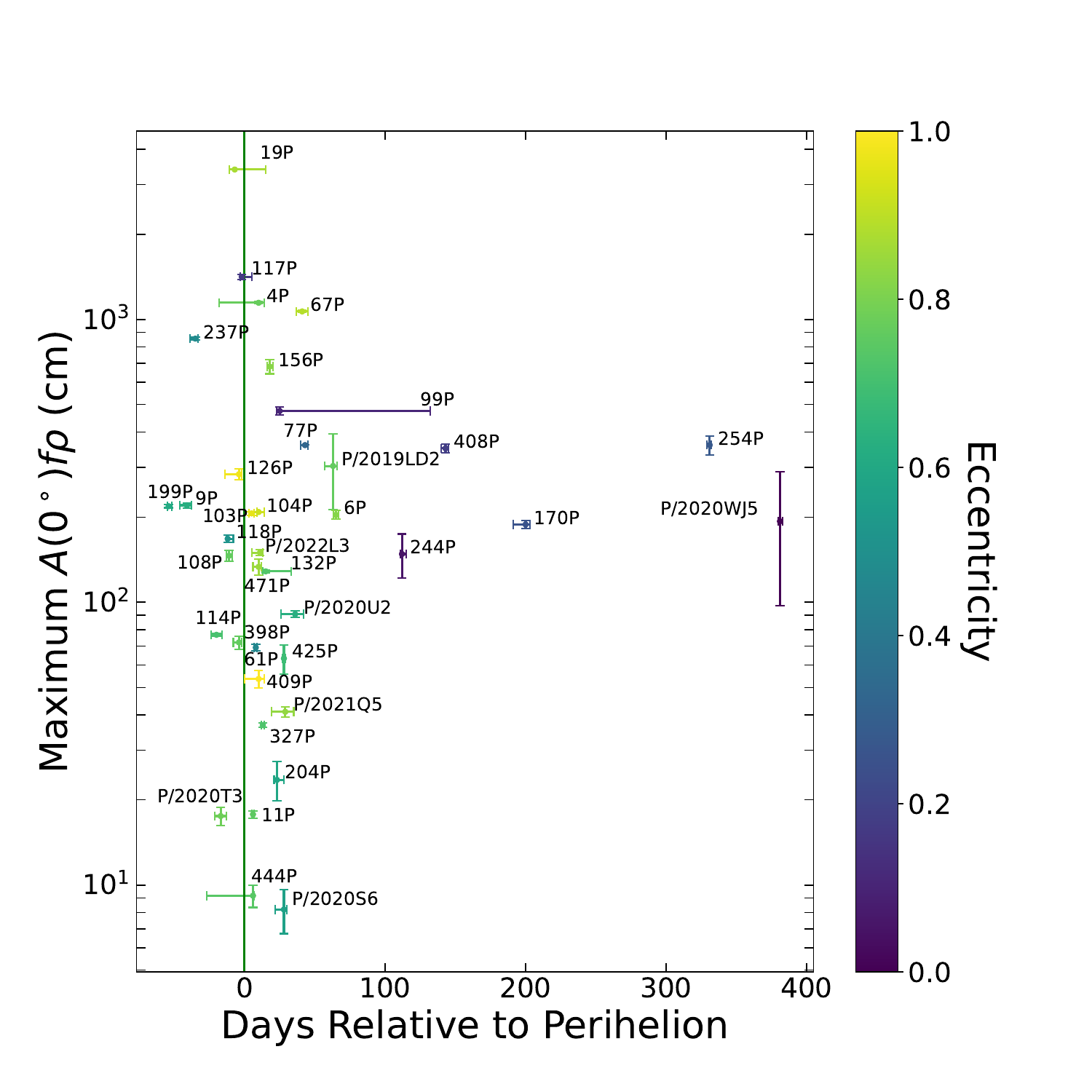}
	\caption{Maximum $A(0^\circ)f\rho$ vs days from perihelion for each JFC is shown. The green vertical line represents the perihelion for each individual comet and the plot is color-coded according to orbital eccentricity.}
	\label{fig:max_activity}
\end{figure}

One may expect the maximum dust production rate in comets to occur at perihelion as this is where the solar flux will be the greatest and therefore the energy available for sublimation will also be at a maximum. Using the same approach as Paper I, we measured where the maximum $A(0^\circ)f\rho$ occurred for each JFC that exhibited a clear maximum. This required an activity curve that was not flat and had a significant coverage of the comet near perihelion. 
Table~\ref{tab:dust_rates} shows the measured maximum $A(0^\circ)f\rho$ and the associated upper limit of dust production rate, $Q_d$, for each JFC in our 2022 and 2023 dataset. We have increased this sample size from 17 JFCs in Paper I to 38 for the combined four-year dataset. Figure~\ref{fig:max_activity} shows the 
maximum activity for each of the 38 JFCs relative to their perihelion. Our findings from Paper I indicated a clear preference for the maximum $A(0^\circ)f\rho$ to occur slightly post-perihelion (within $\sim$ 3 months) and this finding holds true for the expanded dataset as shown in Figure~\ref{fig:max_activity}. 

The reason for this lag is unclear at present.
Previous investigations that have measured the thermal inertia and infrared beaming parameters of cometary nuclei \citep{2009IcarGroussin, 2009A&ALicandro, 2015SciGulkis,  2015A&ASchloerb, 2018A&AMarshall} have found low thermal inertia. Therefore we assume that the comets we have observed also possess a low thermal inertia and therefore rule out a high thermal inertia as the prominent cause of the significant lag in our JFCs reaching maximum $A(0^\circ)f\rho$ relative to perihelion. 

A possible explanation may be the time it takes dust grains to leave the aperture. In Paper I, we showed that the formulae provided by \citet{2002MNRASMa} implied that large 1 mm dust grains would only take $\sim$ 3 days to travel 10,000 km from a 1 km nucleus at $R_h\simeq 1 $ au. At a distance of 2 au, it would take this dust $\sim$ 8 days to leave the aperture, suggesting that this effect could induce a lag time of up to $\sim$8 days in the measured time of maximum activity.

Another contributing factor to extending this `lag time' may be the observational viewing geometry, where the dust remains in the aperture for longer if we observe the comet with our line of sight partly along the direction of the dust tail. 

Figure~\ref{fig:max_activity} indicates that (as found in Paper I) there still remains a number of JFCs reaching their maximum $A(0^\circ)f\rho$ pre-perihelion, a trend we previously attributed to seasonal effects meaning volatile-rich areas of the nculeus are exposed to heating before or after perihelion \citep{2010AJDeSanctis, 2019NatAsTosi}. These seasonal effects can explain why some comets reach their peak $A(0^\circ)f\rho$ just prior to or shortly following perihelion, see \citet{2019AJEhlert, 2024JApAAravind} for examples. Additionally, several comets reach their maximum $A(0^\circ)f\rho$ more than 100 days after perihelion (99P/Kowal 1, 170P/Christensen, 244P/Scotti, 254P/McNaught, 408P/Novichonok-Gerke and P/2020 WJ5 (Lemmon)). These are primarily those with lower eccentricities as we found in Paper I. For these comets, the warming of sub-surface ices may occur more gradually due to their less eccentric orbits.

The range of $A(0^\circ)f\rho$ values in the combined dataset spans well over 2 orders of magnitude, with our lowest activity peak measured in 449P/Leonard (2020 S6) with $A(0^\circ)f\rho$ = 8.2 $\pm$ 0.6 cm. In contrast, the most active JFC we observed in the ATLAS data was 19P/Borrelly, with a peak $A(0^\circ)f\rho$ = 3391 $\pm$ 7 cm. This is very active compared to a typical JFC and the dust activity for this comet is discussed in further detail in Section~\ref{19P_compare}.

For representing the uncertainty in the time of maximum activity we simply used the time span between previous and subsequent measurements either side of maximum in days.
Due to the plateau-like nature of the peak of 99P/Kowal 1, we have instead used the start and end of this plateau to define the uncertainty range. For 99P/Kowal 1, while this JFC exhibits a clear maximum in $A(0^\circ)f\rho$, the peak is less pronounced due to the flatter shape of the activity profile, making it difficult to precisely determine the exact date. In this case, we have identified the highest $A(0^\circ)f\rho$ value within the observed range as the maximum. 


\begin{table*}
\centering
\begin{tabular}{lccccc}
\hline
Comet & Max $A(0^\circ)f\rho$ & $Q_d$ & Days from perihelion & $R_h$ at maximum $A(0^\circ)f\rho$ & $q$\\
 & (cm) & (kg s$^{-1})$ & & (au) & (au)\\
\hline
9P$/$Tempel 1 &   $220.3 \pm 4.5$ &  $\leq$55 &  $-41_{-5}^{+3}$ &             1.60 & 1.54\\
       19P$/$Borrelly &  $3391.2 \pm 7.0$ & $\leq$931 &  $-7_{-4}^{+22}$ &             1.31 & 1.31\\
       61P$/$Shajn-Schaldach &    $69.2 \pm 1.8$ &  $\leq$15 &    $8_{-1}^{+1}$ &             2.13 & 2.13\\
       67P$/$Churyumov-Gerasimenko &  $1067.6 \pm 8.3$ & $\leq$292 &   $41_{-4}^{+4}$ &             1.32 & 1.21\\
       77P$/$Longmore &   $359.6 \pm 0.0$ &  $\leq$73 &   $43_{-3}^{+2}$ &             2.37 & 2.35\\
       99P$/$Kowal 1 &  $474.9 \pm 15.0$ &  $\leq$69 & $25_{-2}^{+107}$ &             4.71 & 4.70\\
      103P$/$Hartley 2 &   $206.3 \pm 3.1$ &  $\leq$63 &    $5_{-2}^{+2}$ &             1.07 & 1.06\\
      104P$/$Kowal 2 &   $208.6 \pm 0.2$ &  $\leq$63 &   $10_{-1}^{+4}$ &             1.08 & 1.07\\
      117P$/$Helin-Roman-Alu 1 & $1416.3 \pm 29.5$ & $\leq$255 &   $-2_{-1}^{+7}$ &             3.04 & 3.05\\
      118P$/$Shoemaker-Levy 4 &   $167.7 \pm 5.2$ &  $\leq$39 &  $-12_{-1}^{+4}$ &             1.83 & 1.83\\
      126P$/$IRAS &  $283.8 \pm 12.0$ &  $\leq$68 &  $-4_{-10}^{+2}$ &             1.71 & 1.71\\
      170P$/$Christensen &   $188.5 \pm 6.1$ &  $\leq$34 &  $200_{-9}^{+3}$ &             3.12 & 2.92\\
      199P$/$Shoemaker 4 &   $218.6 \pm 2.7$ &  $\leq$40 &  $-54_{-1}^{+2}$ &             2.94 & 2.91\\
      204P$/$LINEAR-NEAT &    $23.6 \pm 3.7$ &   $\leq$5 &   $23_{-2}^{+5}$ &             1.85 & 1.83\\
      237P$/$LINEAR &   $855.8 \pm 8.5$ & $\leq$190 &  $-35_{-4}^{+2}$ &             2.01 & 1.99\\
      244P$/$Scotti &  $148.2 \pm 26.3$ &  $\leq$23 &  $112_{-1}^{+3}$ &             3.95 & 3.92\\
      327P$/$Van Ness &    $36.8 \pm 0.7$ &   $\leq$9 &   $13_{-1}^{+1}$ &             1.56 & 1.56\\
      408P$/$Novichonok-Gerke &  $350.2 \pm 11.6$ &  $\leq$59 &  $143_{-3}^{+2}$ &             3.53 & 3.47\\
      444P$/$WISE-PANSTARRS &     $9.2 \pm 0.8$ &   $\leq$2 &   $6_{-33}^{+1}$ &             1.47 & 1.47\\
      471P$/$LINEAR-Fazekas &   $133.4 \pm 8.4$ &  $\leq$29 &   $10_{-4}^{+2}$ &             2.13 & 2.12\\
  P$/$2022 L3 (ATLAS) &   $149.8 \pm 3.5$ &  $\leq$30 &   $11_{-6}^{+2}$ &             2.42 & 2.42\\

\hline
\end{tabular}
\caption{JFC maximum $A(0^\circ)f\rho$ measurements for 2022 and 2023 perihelion comets. $Q_d$ = upper limit of mass loss rate in kg s$^{-1}$.}
\label{tab:dust_rates}
\end{table*}

\subsection{Maximum \texorpdfstring{$Af\rho$}{Af rho} as a function of orbital period} \label{subsec:afrho_vs_orbit}
\begin{figure}[h!]
	\plotone{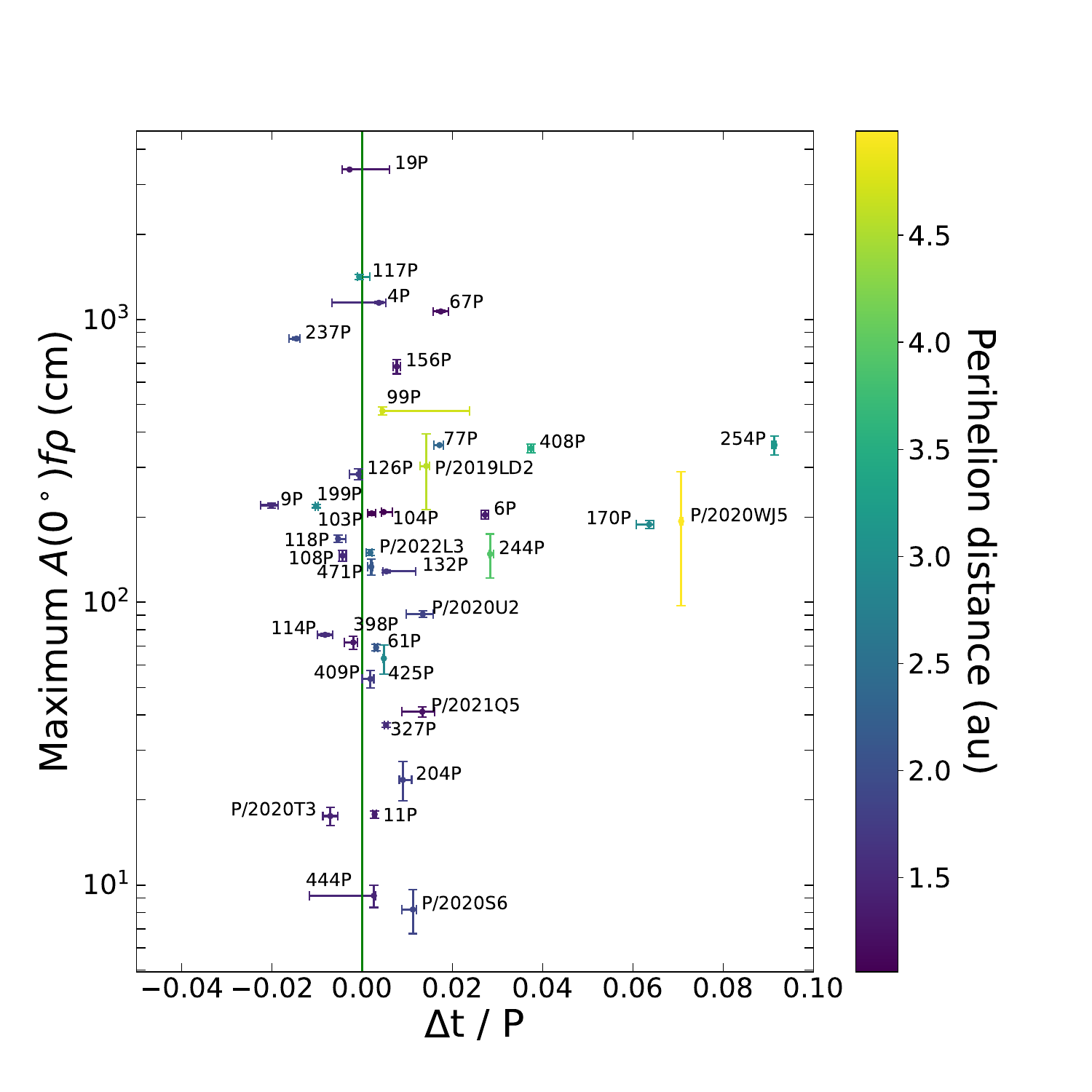}
	\caption{Maximum $A(0^\circ)f\rho$ vs time from perihelion as a function of orbital period for each JFC is shown. The green vertical line represents the perihelion for each individual comet and the plot is color-coded according to perihelion distance.}
	\label{fig:afrho_vs_orbit}
\end{figure}

The relationship between the timing of maximum $A(0^\circ)f\rho$ and the fraction of the orbital period relative to perihelion is shown in Figure~\ref{fig:afrho_vs_orbit}. By considering the orbital period of each JFC rather than focusing solely on the number of days relative to perihelion, we account for the unique orbital characteristics of each comet. To calculate $\Delta t/$P, we simply took the number of days relative to perihelion when maximum $A(0^\circ)f\rho$ was measured and divided by the orbital period in days for each JFC. We found that all comets in our sample—including those that reached their maximum $A(0^\circ)f\rho$ more than a year after perihelion, as shown in Figure~\ref{fig:max_activity}—achieved this maximum within 10\% of their orbital period from perihelion. This demonstrates that, while a year after perihelion may seem like a significant duration, all of the observed JFCs reached their maximum dust production relatively close to perihelion. Additionally, Figure~\ref{fig:afrho_vs_orbit} shows that most of the JFCs that reached their maximum activity significantly later than the rest of the subset (170P/Christensen, 254P/McNaught and P/2020 WJ5 (Lemmon)) tend to have larger perihelion distances, suggesting a connection between perihelion distance and the timing of peak activity. These three JFCs also have lower eccentricities and remain at a similar distance to their perihelion when the maximum $A(0^\circ)f\rho$ is reached, allowing seasonal effects to possibly play a more prominent role in these cases.

Figure~\ref{fig:afrho_vs_orbit} highlights the importance of considering each JFC’s orbit individually, as the dust production rates and activity timelines exhibit significant variability across the JFC population. The absence of JFCs in the top-right region of Figure~\ref{fig:afrho_vs_orbit} shows that such JFCs must be rare, as we would detect them in the ATLAS data. Of course, it is possible that a few JFCs may exist in this region but are not represented in our four-year sample.
The absence of JFCs in the bottom-right region of Figure~\ref{fig:afrho_vs_orbit} is due to observational bias, where such objects exist but remain undetected due to the limitations of our sample size and the sensitivity constraints of the ATLAS telescopes. Comets in this region would be most active at larger distances than perihelion, making their weak peak activity more difficult to detect.

\subsection{Maximum \texorpdfstring{$Af\rho$}{Af rho} and heliocentric distance for four years of ATLAS data} \label{subsec:afp_vs_rh}
\begin{figure}[h!]
	\plotone{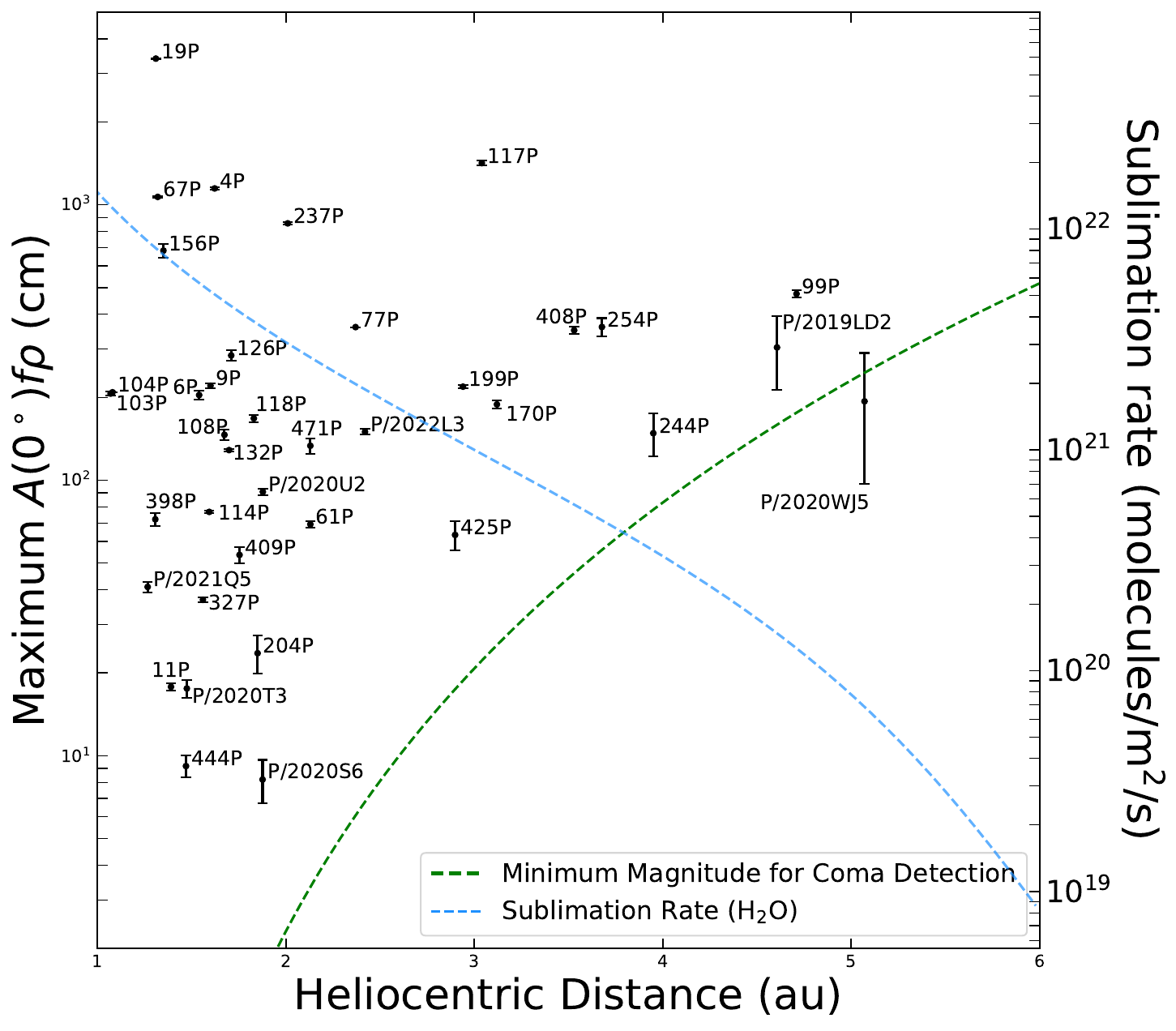}
	\caption{JFC maximum activity as a function of heliocentric distance is shown. The green dashed line indicates the corresponding $A(0^\circ)f\rho$ for $m(o)=18.5$ at opposition, necessary for clear detection of a coma. The absence of measured JFCs below this line is consistent with ATLAS sensitivity. The blue dashed line shows the H${_2}$O sublimation rate as a function of heliocentric distance.}
	\label{fig:afrho_vs_rh2}
\end{figure}

Using the combined dataset,
Figure~\ref{fig:afrho_vs_rh2} illustrates that the majority of the JFCs in our sample reach their maximum $A(0^\circ)f\rho$ within 2.5 au of the Sun. This suggests that water-ice sublimation likely dominated their activity at this time \citep{1985A&AYamamoto, 2015SSRvCochran}. However, several comets, including 408P/Novichonok-Gerke, 254P$/$McNaught, 244P/Scotti, P$/$2019 LD2 (ATLAS), 9P/Kowal 1 and P$/$2020 WJ5 (Lemmon) have a maximum at $R_h\geq 3.4$ au, where the activity may be driven by CO or CO${_2}$. 425P$/$Kowalski, 199P/Shoemaker 4, 117P/Helin-Roman-Alu 1 and 170P/Christensen exhibit a maximum in the transition region at $2.8\leq R_h \leq 3.4$ au, making it unclear if CO/CO${_2}$ or H${_2}$O dominated sublimation is occuring at this time. Figure~\ref{fig:afrho_vs_rh2} also shows the water-ice sublimation rate as a function of heliocentric distance measured in molecules/m$^{2}$/second \citep{2004cometsIIMeech}. We used a bond albedo $A_{B} =0.02$ by calculating $A_{B}=p\times q=0.06 \times0.3$, where $p=0.06$ is the geometric albedo and $q=0.3$ is the phase integral as reported for 19P/Borrelly \citep{2004IcarBuratti}. We also assumed an emissivity of 0.9.

\subsection[AI]{$Af\rho$ activity indices} \label{subsec:AI}
The activity index, $n$, is the power law exponent of how $A(0^\circ)f\rho$ varies with heliocentric distance, $R_{h}$, shown in Equation~\ref{eq:n}.

\begin{equation}\label{eq:n}
A(0^{\circ})f\rho \propto (R_{h})^{n}
\end{equation}

As shown Figure~\ref{fig:327P} (c), $n$ was measured by performing a least squares fit to the pre-perihelion and post-perihelion observations. For several JFCs, $n$ was measured relative to maximum activity rather than perihelion, due to a considerable delay in maximum $A(0^\circ)f\rho$ from perihelion. Additionally, some comets could also only have their activity index determined pre- or post-perihelion due to apparent brightness and/or solar conjunction. We have assumed a power law for the variation of activity as a function of heliocentric distance as that has also been used by many previous studies \citep{1978M&PWhipple, 1997EM&PKidger, 2021PSJWomack, 2024arXivSchleicher}. \citet{2024PSJHolt} observed a significant population of long-period comets and demonstrated that a single power-law fit does not accurately represent the brightening rate across all heliocentric distances. However, for the distances at which ATLAS typically observes JFCs, the use of a single power-law remains appropriate.

As in Paper I, we measured the activity indices of only those JFCs that matched specific criteria. These were JFCs that had six or more observations in our data set taken on separate nights which allowed for a sufficient sample size. The heliocentric distance over which the activity index was measured had to span a range of at least \mbox{$\Delta R_h$ = 0.15 au}. Finally, the data quality for each JFC was evaluated based on the effective signal to noise. The signal to noise was determined by dividing each $A(0^\circ)f\rho$ measurement by its associated uncertainty. We generally found an average signal to noise $>7$ allowed an acceptable measurement of $n$. 
In our 2022 and 2023 dataset we have measured the activity indices for 37 JFCs, listed in Table~\ref{tab:AI}. 

The activity indices ranged from $-12.5$ to $5.5$ pre-perihelion and $-8.0$ to $1.3$ post-perihelion. The majority of the JFCs in Table~\ref{tab:AI} possess negative activity indices as should be expected for gas sublimation driven activity, but there are a few that possess positive values either in the pre- or post-perihelion samples. These were mostly where the variation in $A(0^\circ)f\rho$ was smaller than the measurement uncertainties, leading to apparently flat activity curves. In the case of 285P/LINEAR, this comet has a highly positive activity index of $n$ = 5.5 $\pm$ 0.3 after we removed the outburst for our measurement. This positive $n$ value is due to the comet's decreasing activity in the pre-perihelion phase. However, because this activity index is derived over a relatively short span of $\sim$ 80 days and $\sim$ 0.35 au, this activity index should be interpreted with caution as such a limited window may not fully represent the comet's long-term activity behaviour.

\begin{figure}[h!]
	\plotone{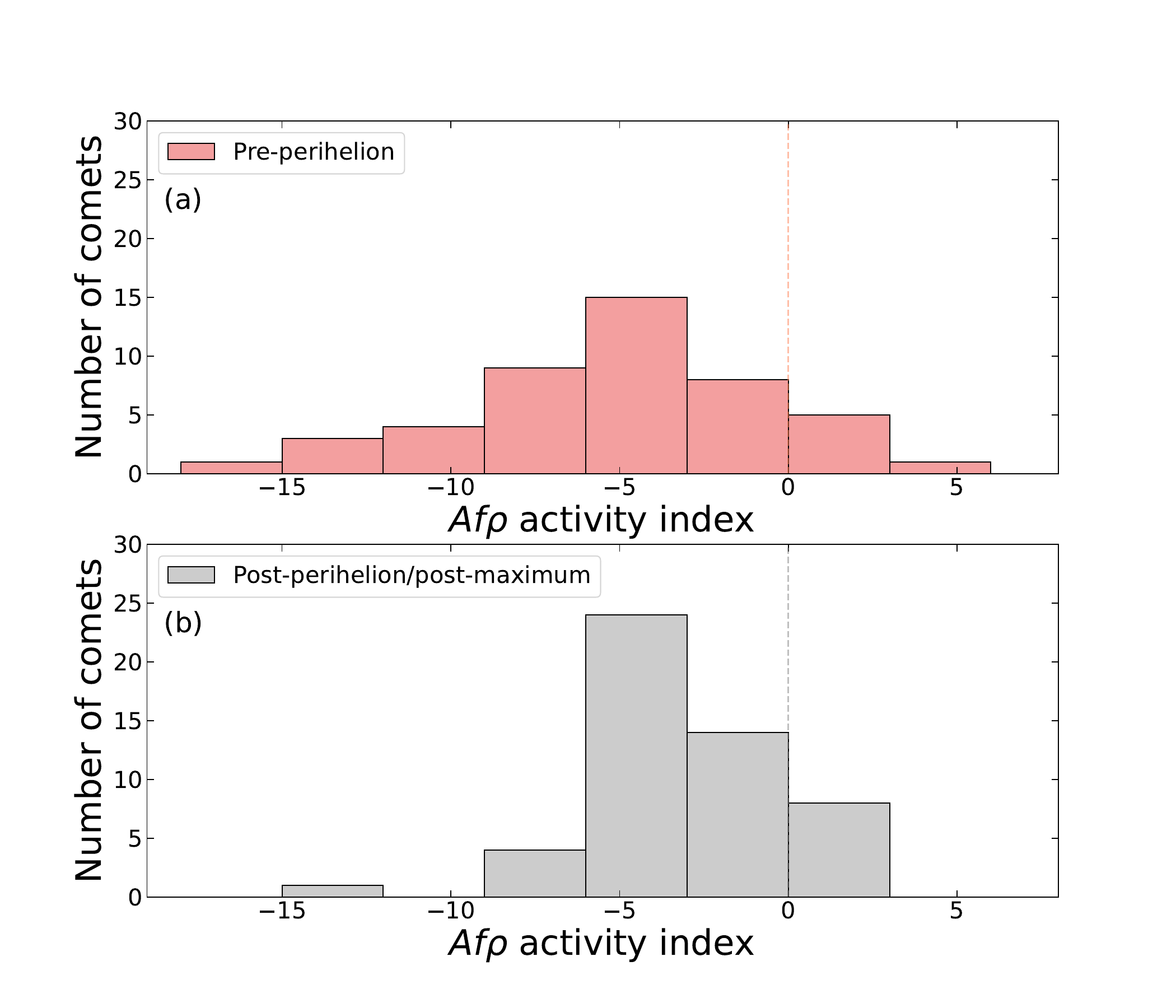}
	\caption{The top panel, (a) shows the activity index, $n$, for the JFC subset pre-perihelion. Plot (b) shows $n$ for JFCs post-perihelion/post-max. Binned to $\Delta n = 3$.}
	\label{fig:AI_histogram}
\end{figure}

The average activity index for the four-year sample was $-5.2 \pm 4.5$ pre-perihelion and  $-3.2 \pm 2.7$ post-perihelion. Hence, the average rate of change of dust production as a function of heliocentric distance is greater pre-perihelion than post-perihelion. The distribution of activity indices is shown in Figure~\ref{fig:AI_histogram}.
Overall, the measured activity indices pre-perihelion exhibit a much broader range than those post-perihelion, with greater variation among the number of JFCs sharing a similar activity index. Conversely, the activity indices post-perihelion display a narrower range, with the majority falling within $-6 \leq n \leq -3$ in our four-year sample. 

The reason for the large range of pre-perihelion values is likely due to the significant differences in the physical properties of JFC nuclei, including shape, size, fractional active areas, amount of volatiles, and different near-surface volatile distributions. Activity driven by gas sublimation would predict an activity index of $n=-2$ due to the dependence on sublimation which falls off as the inverse square of heliocentric distance. An activity index of $n=-2$ only theoretically applies to cometary nuclei at small heliocentric distances where all of the absorbed solar energy goes into sublimation and there is no heat transferred to the subsurface. As the heliocentric distance increases, $n$ depends on the relative volatility of the grains and nucleus as well as the initial size of the grains \citep{1980M&PAHearn}. 

If we consider dust fragmentation, the grains that are larger that $\sim$10 $\mu$m \citep{2019A&AMannel} and are ejected at lower velocities by the gas-drag force, can fragment in the coma to form smaller dust grains. If we assume a dust grain that fragments into two smaller dust grains of equal size, the visible cross-sectional area of the dust would increase by a factor of 1.3. This repeated fragmentation can lead to an exponential brightening as seen in 17P/Holmes \citep{2010MNRASHsieh}. As the comet approaches perihelion and the fragmentation rate increases, the total brightness of the comet will also increase. If significant dust fragmentation occurred in the coma of any of our JFCs, we may expect a steeper activity index than $n=-2$.

If we consider crust insulation \citep{1989ESASPThiel} occurring on a given JFC nucleus, this may also steepen the activity slope. Dust grains that are too large to be ejected will remain on the nucleus and form an insulating crust that can cover the ices \citep{1996EM&PJewitt, 2024A&AMastropietro}. As the comet approaches perihelion and the nucleus is exposed to increasing amounts of solar energy, and therefore typically a temperature increase, the dust mantle may be partially or fully lost, exposing fresh volatile ices that readily sublimate.

It is challenging to assess the extent to which dust fallback \citep{2020FrPMarschall} influences the activity index as several factors contribute. As the comet nucleus approaches the Sun, the increasing solar insolation enhances the overall sublimation, creating a larger gas pressure gradient capable of ejecting progressively larger dust grains which travel at lower velocities. Alternatively, at smaller heliocentric distances, the higher gas pressure gradient results in greater ejection velocities, allowing larger grains to achieve escape velocity and avoid falling back onto the nucleus. It is therefore difficult to determine the net dust fallback and how much it contributes to the overall activity index. Near perihelion, larger dust particles are ejected from the nucleus at lower velocities. These slower speeds result in the particles taking longer to vacate the aperture, resulting in a shallower activity index, $n$, post-perihelion.

\begin{table*}
\centering
\caption{JFC $A(0^\circ)f\rho$ and activity index $n$ pre- and post-perihelion for the 2022 and 2023 perihelion JFCs. $A(0^\circ)f\rho$ pre and post-perihelion were interpolated and extrapolated to 2au.}
\begin{tabular}{lcccc}
\hline
Name & $A(0^\circ)f\rho$ pre-perihelion & $A(0^\circ)f\rho$ post-perihelion &  $n$ pre-perihelion & $n$ post-perihelion \\
\hline
9P/Tempel 1 &        - &      $93.1 \pm 1.9$ &   - & $-2.0 \pm 0.04$ \\
             19P/Borrelly &      $209.9 \pm 8.1$ &    $454.7 \pm 14.1$ &  $-6.3 \pm 0.1$ & $-4.3 \pm 0.1$ \\
                22P/Kopff &     $131.8 \pm 34.7$ &     $203.1 \pm 7.3$ &  $-2.2 \pm 0.3$ & $-3.5 \pm 0.1$ \\
           44P/Reinmuth 2 &     $314.5 \pm 50.6$ &      $66.2 \pm 6.7$ &  $-4.9 \pm 0.2$ & $-1.0 \pm 0.2$ \\
      61P/Shajn-Schaldach &      $82.3 \pm 11.5$ &       - &  $-3.5 \pm 0.2$ &  - \\
67P/Churyumov-Gerasimenko &      $148.8 \pm 3.1$ &     $258.0 \pm 4.7$ &  $-3.2 \pm 0.04$ & $-3.4 \pm 0.03$ \\
                71P/Clark &      $63.6 \pm 18.4$ &    $140.0 \pm 12.1$ &  $-3.1 \pm 0.4$ & $-4.7 \pm 0.1$ \\
             77P/Longmore &     $680.1 \pm 54.0$ &    $623.2 \pm 31.9$ &  $-5.5 \pm 0.1$ & $-4.0 \pm 0.1$ \\
               81P/Wild 2 &     $888.8 \pm 83.0$ &    $408.9 \pm 12.8$ &  $-5.4 \pm 0.1$ & $-2.7 \pm 0.04$ \\
           96P/Machholz 1 &       $20.7 \pm 7.0$ &      $54.1 \pm 2.9$ &   $0.6 \pm 0.5$ & $-1.8 \pm 0.1$ \\
   97P/Metcalf-Brewington &     $117.1 \pm 65.0$ &       - &  $-5.0 \pm 0.8$ &  - \\
              99P/Kowal 1 & $14003.3 \pm 9334.8$ & $8212.6 \pm 2735.6$ &  $-4.3 \pm 0.7$ & $-3.5 \pm 0.3$ \\
           103P/Hartley 2 &       $20.7 \pm 1.5$ &      $51.0 \pm 1.1$ &  $-2.0 \pm 0.2$ & $-1.1 \pm 0.04$ \\
             104P/Kowal 2 &        $2.3 \pm 0.3$ &      $81.4 \pm 1.7$ &  $-5.2 \pm 0.3$ & $-1.7 \pm 0.04$ \\
            113P/Spitaler &     $235.3 \pm 96.6$ &      $31.9 \pm 6.2$ &  $-8.3 \pm 0.6$ &  $0.1 \pm 0.3$ \\
   117P/Helin-Roman-Alu 1 &   $9972.5 \pm 579.8$ & $12869.8 \pm 801.0$ &  $-4.9 \pm 0.1$ & $-5.7 \pm 0.1$ \\
    118P/Shoemaker-Levy 4 &       $84.6 \pm 9.8$ &     $87.7 \pm 11.1$ &  $-7.5 \pm 0.2$ & $-4.5 \pm 0.2$ \\
      119P/Parker-Hartley &    $592.0 \pm 101.4$ &    $533.1 \pm 24.9$ &  $-5.4 \pm 0.2$ & $-3.4 \pm 0.1$ \\
                126P/IRAS &        - &     $160.3 \pm 5.7$ &   - & $-3.3 \pm 0.1$ \\
      152P/Helin-Lawrence & $20583.2 \pm 8648.4$ &      $85.9 \pm 5.2$ & $-10.3 \pm 0.5$ &  $0.6 \pm 0.1$ \\
         170P/Christensen &   $1737.1 \pm 803.5$ &   $804.9 \pm 236.4$ &  $-7.8 \pm 0.6$ & $-3.4 \pm 0.4$ \\
                180P/NEAT &    $888.1 \pm 344.4$ &       - & $-10.5 \pm 0.6$ &  - \\
         199P/Shoemaker 4 &    $376.0 \pm 149.6$ &                    - &  $-1.7 \pm 0.5$ &  - \\
         204P/LINEAR-NEAT &        - &      $20.3 \pm 3.4$ &   - & $-0.9 \pm 0.3$ \\
              237P/LINEAR &    $833.0 \pm 144.2$ &    $552.2 \pm 28.4$ & $-12.5 \pm 0.3$ & $-8.0 \pm 0.1$ \\
              244P/Scotti &    $338.0 \pm 154.6$ & $4316.3 \pm 2159.4$ &  $-1.6 \pm 0.5$ & $-5.0 \pm 0.6$ \\
               263P/Gibbs &        $6.3 \pm 1.5$ &       $5.5 \pm 2.0$ &   $2.0 \pm 0.1$ &  $1.3 \pm 0.8$ \\
    274P/Tombaugh-Tenagra &     $223.2 \pm 78.5$ &       - &  $-8.5 \pm 0.5$ &  - \\
              285P/LINEAR &     $102.4 \pm 16.5$ &       - &   $5.5 \pm 0.3$ &  - \\
            327P/Van Ness &        $1.4 \pm 0.3$ &      $11.7 \pm 0.5$ & $-11.9 \pm 0.4$ & $-4.3 \pm 0.1$ \\
           364P/PANSTARRS &        $2.8 \pm 0.6$ &       $3.5 \pm 0.4$ &  $-1.6 \pm 0.5$ & $-3.1 \pm 0.3$ \\
    408P/Novichonok-Gerke &  $7751.1 \pm 1272.4$ &   $688.5 \pm 131.9$ &  $-6.5 \pm 0.19$ & $-1.9 \pm 0.2$ \\
      444P/WISE-PANSTARRS &        $2.1 \pm 0.6$ &                    - &  $-3.4 \pm 0.6$ &  - \\
            459P/Catalina &        - &      $21.4 \pm 2.7$ &   - &  $0.4 \pm 0.3$ \\
      471P/LINEAR-Fazekas &     $227.5 \pm 39.4$ &       - &  $-8.6 \pm 0.3$ &  - \\
        P/2022 L3 (ATLAS) &     $168.8 \pm 17.0$ &    $264.0 \pm 35.9$ &  $-0.8 \pm 0.2$ & $-3.0 \pm 0.21$\\
\hline
\end{tabular}
\label{tab:AI}
\end{table*}

To determine the statistical significance of the results shown in Figure~\ref{fig:AI_histogram}, we performed a two sample Kolmogorov–Smirnov test (K-S test) on our entire four-year dataset, the null hypothesis being that the pre- and post-perihelion values of $n$ are sampled from the same distribution.
We obtained values of d $=0.29$ and p $=0.015$. Our p-value is less than the chosen significance level of 0.05 and therefore we reject the null hypothesis that the two samples are drawn from the same distribution.  We have now confirmed the two samples differ significantly when using the larger sample size of 56 JFCs than in Paper I with only 20. Therefore, we conclude that JFCs typically exhibit a shallower activity index post-perihelion than pre-perihelion.

\subsection{Dust production of JFCs at 2 au} \label{subsec:2au}
\begin{figure}[h!]
	\plotone{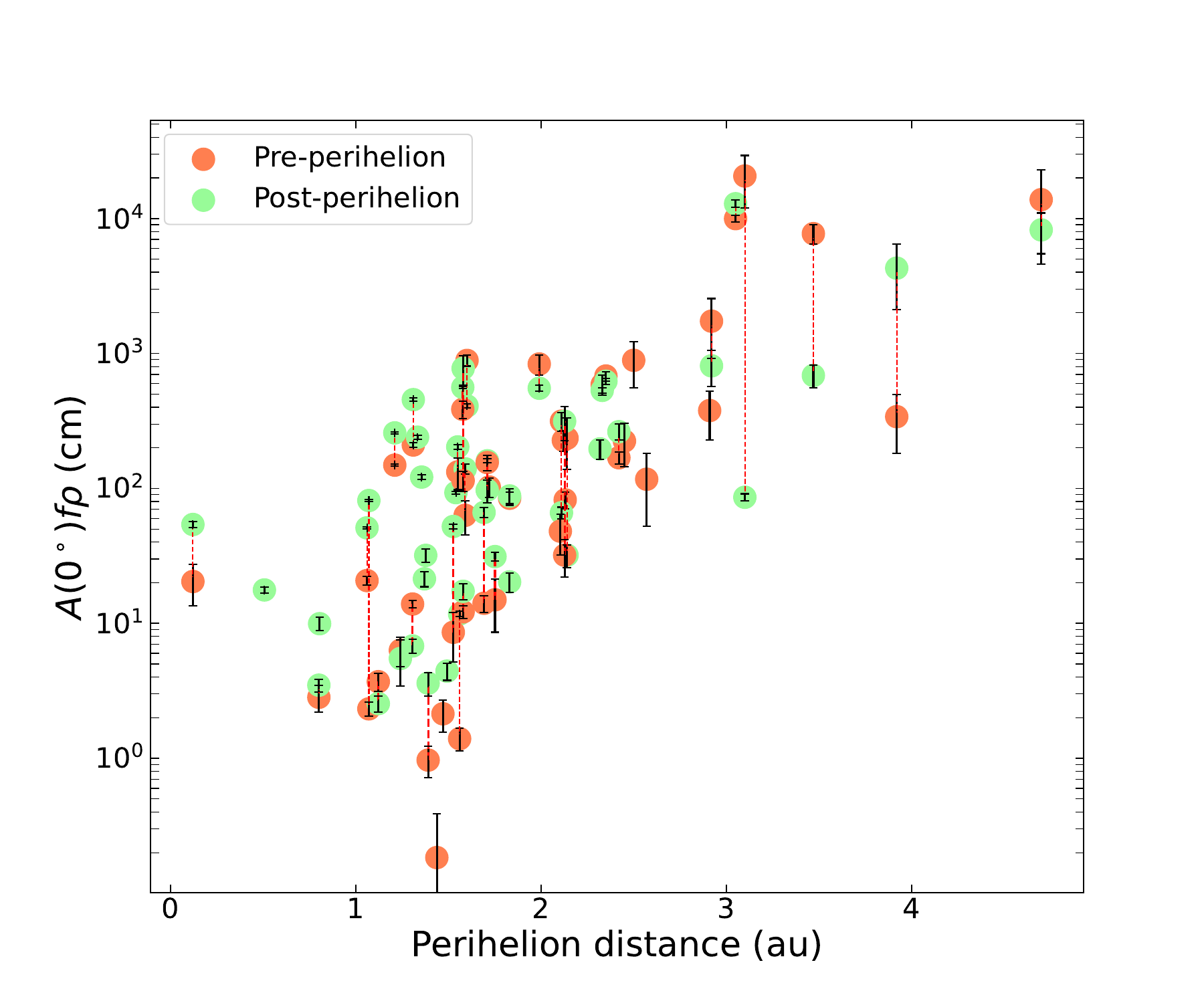}
	\caption{$A(0^\circ)f\rho$ projected to 2 au against the perihelion distance for each JFC that has a measurable $n$ in the 4 year ATLAS sample. Orange points show the JFCs projected $A(0^\circ)f\rho$ at 2 au and green points show the projected $A(0^\circ)f\rho$ at 2 au post-perihelion. The red dashed line links the observations of the same JFC pre-perihelion to post-perihelion, where both measurements were made.}
	\label{fig:2au_extrapolate}
\end{figure}

To measure the intrinsic variation in $A(0^\circ)f\rho$ within the JFC population, we have used the measured activity indices to interpolate and extrapolate the observed values of $A(0^\circ)f\rho$ to a common distance of 2 au. Most of the JFCs in our sample were observed at \mbox{$<$ 3 au} and water sublimation should dominate. 
 In paper I, we concluded that there no discernible correlation between JFC activity and perihelion distance. The combined dataset presented in this paper allows us to further investigate whether the initial lack of correlation was simply due to limited data and are shown in Figure~\ref{fig:2au_extrapolate}. We also observed a sensitivity bias in the bottom-right region of Figure~\ref{fig:2au_extrapolate}, where comets with low $A(0^\circ)f\rho$ values at larger perihelion distances fall below the ATLAS detection threshold.

We find that there is still no observable correlation between the intrinsic activity of the JFCs and their perihelion distance. This lack of correlation suggests that the dominant influence on the dust activity is likely the nucleus itself. JFC nuclei exhibit significant variability in shape and size, have different fractional active areas, and possess different distributions of volatile material for sublimation. Each of these factors could contribute to the activity of the JFC by varying amounts and therefore isolating a single causal factor is a challenging task.

All of the extrapolated $A(0^\circ)f\rho$ values at 2 au for comets with perihelion distances greater than 2 au should be treated with caution. These comets must exhibit a relatively high level of activity to be detectable at such large distances. Consequently, when their activity is extrapolated inwards to 2 au, the resulting $A(0^\circ)f\rho$ values tend to be disproportionately higher than the bulk of closer JFCs. This effect is particularly evident in the top-right region of Figure~\ref{fig:2au_extrapolate}, where the projected values are much larger, likely due to the over-representation of highly active JFCs in this subset. This skew effect may be minimised by considering only JFCs with perihelion distances less than 2 au, ensuring a direct comparison among similar comets. This would reduce observational bias toward JFCs with larger perihelion distances, which tend to have disproportionately higher projected $A(0^\circ)f\rho$ values at 2 au. However, we have included all JFCs with measurable activity indices for completeness and to maintain consistency with the JFCs analysed in Paper I.

The JFC that showed the largest change in $A(0^\circ)f\rho$ from pre-perihelion to post-perihelion was 152P/Helin-Lawrence. Its activity appeared to decrease drastically from a projected pre-perihelion value of $A(0^\circ)f\rho$ = (2 $\pm$ 1)$\times 10^{4}$ cm to $A(0^\circ)f\rho$ = 86 $\pm$ 5 cm. However, this is highly atypical for a JFC to have such a large measured $A(0^\circ)f\rho$ and can be explained by several factors. First, 152P/Helin-Lawrence has a perihelion distance of $q=3.1$ au, and thus we are projecting $A(0^\circ)f\rho$ values to a distance 1.1 au closer than the actual perihelion. It is possible that this JFC has never undergone water-ice driven sublimation, and thus, the projection may not reflect its true dust activity. Additionally, the activity index for 152P/Helin-Lawrence, while meeting the criteria established in Section~\ref{subsec:AI}, may not be suitable for projecting its behaviour. The pre-perihelion activity of 152P/Helin-Lawrence begins to flatten, indicating that a constant activity index is not appropriate for this JFC. This flattening suggests that this method of projection overestimates the comet's activity at 2 au.

For the majority of the JFCs shown in Figure~\ref{fig:2au_extrapolate}, which have both pre- and post-perihelion $A(0^\circ)f\rho$ projected measurements, the change in $A(0^\circ)f\rho$ is typically less than an order of magnitude. Excluding outbursts, these relatively modest changes in activity more accurately reflect the general pattern expected from JFCs, where dust production rates evolve smoothly and gradually as the comets approach and recede from perihelion. This is also consistent with models of cometary activity, such as that presented in \citet{2002EM&PPrialnik}, which show similarly smooth and moderate variations in activity across perihelion.

\section{Discussion}
\subsection{Nuclear radii upper limit estimates} \label{subsec:n_detects}
\begin{figure}[h!]
	\plotone{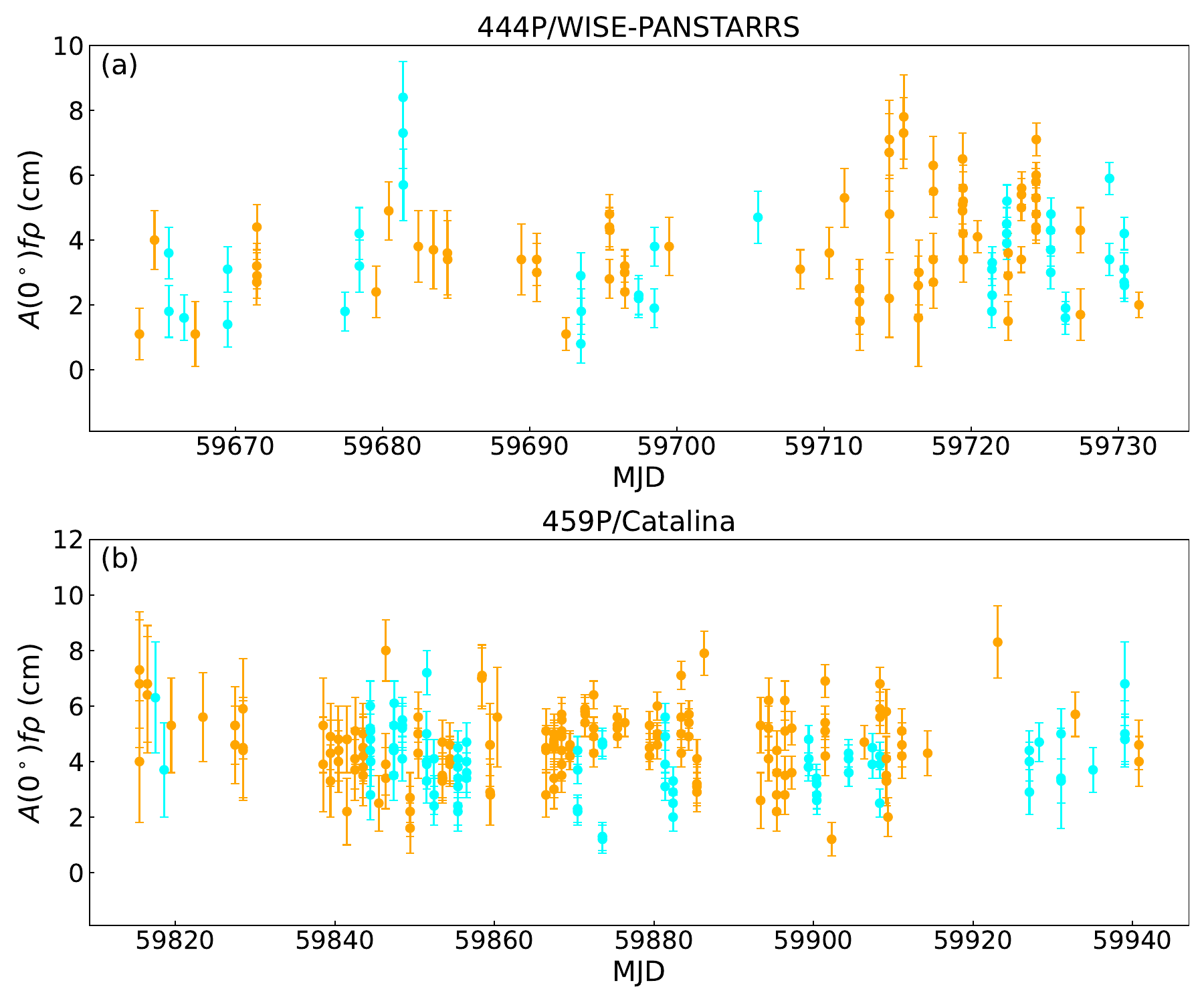}
	\caption{$A(0^\circ)f\rho$ for comets 444P/WISE-PANSTARRS and 459P/Catalina during pre-perihelion dates when we obtained constant $A(0^\circ)f\rho$ measurements within the measurement uncertainties. 444P/WISE-PANSTARRS is shown from $R_{h}$ = 1.84 au -- 1.53 au, and 459P/Catalina is shown from $R_{h}$ = 2.27 au -- 1.50 au.} Orange and cyan points on the plots represent the ATLAS \emph{o-}band and \emph{c-}band.
	\label{fig:Rn2}
\end{figure}

Two of the JFCs detected in our 2022 and 2023 JFC sample, 444P/WISE-PANSTARRS and 459P/Catalina, showed no visible coma for $>60$ days. Figure~\ref{fig:Rn2} shows the corresponding pre-perihelion measurements for these comets, where they exhibit no appreciable change in $A(0^\circ)f\rho$.

This is consistent with photometry dominated by a direct detection of the nucleus. Using the same approach as Paper I, we used these data to determine upper limits to their nucleus radii, assuming a geometric albedo of $0.04$.
444P/WISE-PANSTARRS has $R_n$ $\leq$ $1.5\pm0.2$ km, and 459P/Catalina has $R_n$ $\leq$ $1.7\pm0.1$ km. Both JFCs have no previous nuclear radii measurements reported in the literature to the best of our knowledge. The $R_n$ of both JFCs calculated in this study are consistent with the sizes of JFCs found in previous works, see \citet{2011MNRASSnodgrass, 2023arXivKnight}.


\subsection[Comparison]{Comparison of JFC $Af\rho$ values with previous studies}
We compared the dust production rates found in this work to previous measurements available in the literature. From our sample of 74 JFCs reaching perihelion in 2022 and 2023, only a small number have had their $Af\rho$ measurements previously reported. 

Nine of our comets appear in a study by \citet{2009IcarFink} that used CCD spectroscopy to measure $Af\rho$ values. The $Af\rho$ values calculated in that study were all corrected to a phase angle of 40\textdegree\ using the dust scattering phase function from \citet{1981ESASPDivine}. We corrected these measurements to a phase angle of 0\textdegree\ using the \citet{2010Schleicher} phase function to allow comparison to our measurements. This approach has been applied uniformly across all relevant cases. Most of our measurements disagree with those of \citet{2009IcarFink}. Spectroscopic measurements of $Af\rho$ often differ from measurements using circular apertures as discussed in \citet{2019MNRASHyland}, probably because of the non-spherical symmetry in the coma which gives a brightness profile that deviates from $\rho^{-1}$. We have compared our measurements nevertheless, as it is unclear where any discrepancy arises for a particular comet.

\subsubsection{9P/Tempel 1}
9P/Tempel 1 was the target of the \emph{Deep Impact} mission in 2005 \citep{2005SciAHearn}, which was the first mission to impact a cometary nucleus. 9P/Tempel was then subsequently visited by the re-purposed \emph{Stardust NExT} spacecraft in 2011 \citep{2013IcarVeverka} after its original visit to 81P/Wild 2 being the first comet to be visited by multiple different spacecraft.
As 9P/Tempel 1 is one of the most studied comets to date, there are $Af\rho$ measurements of this comet dating back to the 1980's. This comet has a relatively short orbital period of 5.6 years, and \citet{2007IcarMilani} found 9P/Tempel 1 displayed a similar behaviour between 1983 and 2005. Previous studies containing $Af\rho$ measurements from the 80's and 90's were by \cite{2007IcarSchleicher,2011IcarMeech,2007IcarMilani,2009IcarFink}.

\citet{2007IcarMilani} discuss the time between maximum activity and perihelion for the 2004--2005 apparition, similar to Section~\ref{subsec:afrho_vs_days} in this work. Their measured $Af\rho$ plateaus around 420 cm and is `more flat' and spans between 85 and 4 days before perihelion. In our ATLAS data, shown in Table~\ref{tab:dust_rates}, we measured 9P/Tempel 1 to reach the maximum activity $-41_{-5}^{+3}$ days before perihelion which is consistent with this study. The reason for the maximum activity pre-perihelion is possibly due to strong seasonal effects associated with 9P/Tempel 1 as discussed by \citet{2007IcarSchleicher}. 

\cite{2013IcarFarnham} measured $A(79^\circ)f\rho$ = 43 $\pm$ 6 cm on 2011 February 9 at $R_{h}$ = 1.54 au. Using the \citet{2010Schleicher} dust phase function to correct to a phase angle of $\alpha$ = 0\textdegree, resulting in a measured $A(0^\circ)f\rho$ = 99 $\pm$ 14 cm on approach to the target from the \emph{Stardust-NExT NAVCAM} images. This apparition was two orbits previous to the measurements we have made with ATLAS of significantly more activity with $A(0^\circ)f\rho$ = 179 $\pm$ 4 cm at $R_{h}$ 1.55 au. The authors note that NAVCAM yields broadband measurements from 475–925 nm and therefore their measurement may be contaminated by the emission of various molecular fluorescent bands (predominantly $C_{2}$ and $NH_{2}$). The discrepancy between our two measurements may also arise from the smaller apertures the authors used, which ranged from 3300--6500 km.

\subsubsection{19P/Borrelly} \label{19P_compare}
\citet{2009IcarFink} observed an increase in dust activity from $A(0^\circ)f\rho$ = 944 cm at $R_{h}$ = 1.51 au on 1994 September 6 to $A(0^\circ)f\rho$ = 1383 cm at $R_{h}$ = 1.41 au on 1994 December 3, either side of perihelion in November 1994. We measured $A(0^\circ)f\rho$ = 94 $\pm$ 3 at $R_{h}$ = 1.53 au on 2021 August 7 and $A(0^\circ)f\rho$ = 1939 $\pm$ 3 cm at $R_{h}$ = 1.42 au post-perihelion on 2022 March 20. Our measurements are not consistent with the earlier reported values from the 1994 apparition. We note that the comet exhibited significantly higher activity post-perihelion and is the most active JFC in our entire four-year study. Although our measurements differ from the previous study, 19P/Borrelly shows a wide range of activity, from near-inactive to a maximum $A(0^\circ)f\rho$ = 3391 $\pm$ 7 cm as shown in Table~\ref{tab:dust_rates}. Despite the discrepancies, the measurements by \citet{2009IcarFink} fall within the range of our observations.

\subsubsection{22P/Kopff}
\citet{2009IcarFink} measured the $Af\rho$ of 22P/Kopff at two different epochs and measured a decrease of $A(0^\circ)f\rho$ = 396 cm at $R_{h}$ = 2.14 au on 1990 June 20, to $A(0^\circ)f\rho$ = 228 cm at $R_{h}$ = 2.81 au on 1990 October 11. We measured $A(0^\circ)f\rho$ = 158 $\pm$ 2 cm at $R_{h}$ = 2.15 au on 2022 August 19, decreasing to $A(0^\circ)f\rho$ = 62 $\pm$ 10 cm at $R_{h}$ = 2.81 au on 2022 December 7. Both of our $A(0^\circ)f\rho$ measurements are smaller and the absolute decrease is also smaller. 

At a second apparition, \citet{2009IcarFink} measured another decrease in $Af\rho$ from 22P/Kopff pre-perihelion of $A(0^\circ)f\rho$ = 923 cm at $R_{h}$ = 1.65 au on 1996 May 16, to $A(0^\circ)f\rho$ = 791 cm at $R_{h}$ = 1.59 au on 1996 June 16. We only have activity measurements at $R_{h}$ = 1.57 au of $A(0^\circ)f\rho$ = 474 $\pm$ 3 cm at on 2022 April 12 due to solar conjunction, but this is significantly lower than that of the comparison study. The discrepancy could be due to the comet being less active on this orbit than the 1990 apparition, or the general discrepancy between spectroscopy and photometry mentioned above.

\subsubsection{44P/Reinmuth 2}
\citet{2009A&ALamy} used HST snapshot observations in the F675W filter to measure $A(0^\circ)f\rho$ = 105 $\pm$ 1 cm for comet 44P/Reinmuth 2 inbound on its orbit in June 2000 at $R_{h}$ = 2.7 au. We obtained $A(0^\circ)f\rho$ = 50 $\pm$ 6 cm at $R_{h}$ = 2.7 au 2021 August 31. Although our measured $A(0^\circ)f\rho$ is lower than the comparison study, this may be expected as our nightly variation in $A(0^\circ)f\rho$ is substantial. The authors also used a much smaller aperture of 400 km diameter.

\citet{2001A&ALowry} determined an $A(0^\circ)f\rho$ upper limit of $\leq$ 3.3 cm using the 4.2 m William Herschel Telescope in December 1998 at an $R_h$ = 4.7 au inbound. We do not have measurements of 44P/Reinmuth 2 beyond $R_{h}$ = 3.2 au in this study so comparison is not possible.

\subsubsection{67P/Churyumov-Gerasimenko}
67P/Churyumov-Gerasimenko was the target of the \emph{ESA} \emph{Rosetta} mission \citep{2007Glassmeier, 2015SciSierks}. This was the first mission to rendezvous with a comet and remain with the comet throughout a large portion of its orbit. The \emph{Rosetta} mission was also the first to deploy a lander (\emph{Philae}) onto a cometary nucleus in November 2014.

\citet{2013A&ASnodgrass} measured $A(0^\circ)f\rho$ within 10,000 km in the weeks after perihelion of $Af\rho$ $\sim$ 1000 cm during the 2007/2008 apparition. The authors measured $Af\rho$ $\sim$ 100 cm at 2 au, and $Af\rho \sim 50$ cm at 3 au. Our measurements of 67P at perihelion, 2 au and $\sim 3$ au are consistent with this comparison study. The consistency between the two studies shows that the comet was roughly the same level of activity on this apparition as the 2007/2008 apparition. \citet{2013A&ASnodgrass} also measured an activity index of $n = -3.2$ pre-perihelion, and $n = -3.4$ post-perihelion. For the 2021/2022 apparition, we measured $n -3.18 \pm 0.04$ pre-perihelion, and $n = -3.35 \pm 0.03$ post-perihelion, consistent with 2007/2008.

\citet{2017RSPSnodgrass} presented a summary of the campaign of remote observations that supported the \emph{Rosetta} mission from telescopes across the globe (and in space) from before \emph{Rosetta’s} arrival until nearly the end of the mission in September 2016. Again, the authors found that the peak in activity was around $Af\rho \sim 1000$ cm. This is consistent with our maximum $Af\rho$ measurements for 67P during its most recent orbit, see Table~\ref{tab:dust_rates}. \citet{2017RSPSnodgrass} noted that 67P is a relatively `well-behaved’ comet, typical of JFCs and with an activity pattern that repeats from orbit to orbit.

\cite{2016MNRASBoehnhardt} also observed 67P on 51 nights between 2015 August 22 and 2016 May 9 from $R_{h} =$ 1.2 to 3 au. Using a 10,000 km aperture and SDSS--\emph{r\textquotesingle} filter, the authors measured a decrease in $A(0^\circ)f\rho$ from 840 cm to 46 cm during this period. At 1.2 au, we measured $A(0^\circ)f\rho$ = 881 $\pm$ 1 cm and at 2.8 au (the maximum $R_{h}$ we observed post-perihelion), we measured $A(0^\circ)f\rho = 61 \pm 15$ cm. Although our results are not consistent within 1$\sigma$, the results and overall evolution of $Af\rho$ between the two studies is similar. Finally, \cite{2017MNRASKnight} observed 67P in 2014/2016 using a mixture of SDSS and Cousins--R filters. The measured fluxes were corrected using the Marcus-Schleicher phase function, and they found a peak R-band $A(0^\circ)f\rho \simeq 1700$ cm post-perihelion, of similar order but larger than the studies cited above.

\begin{figure}[h!]
    \centering
    \epsscale{1.25}
    \plotone{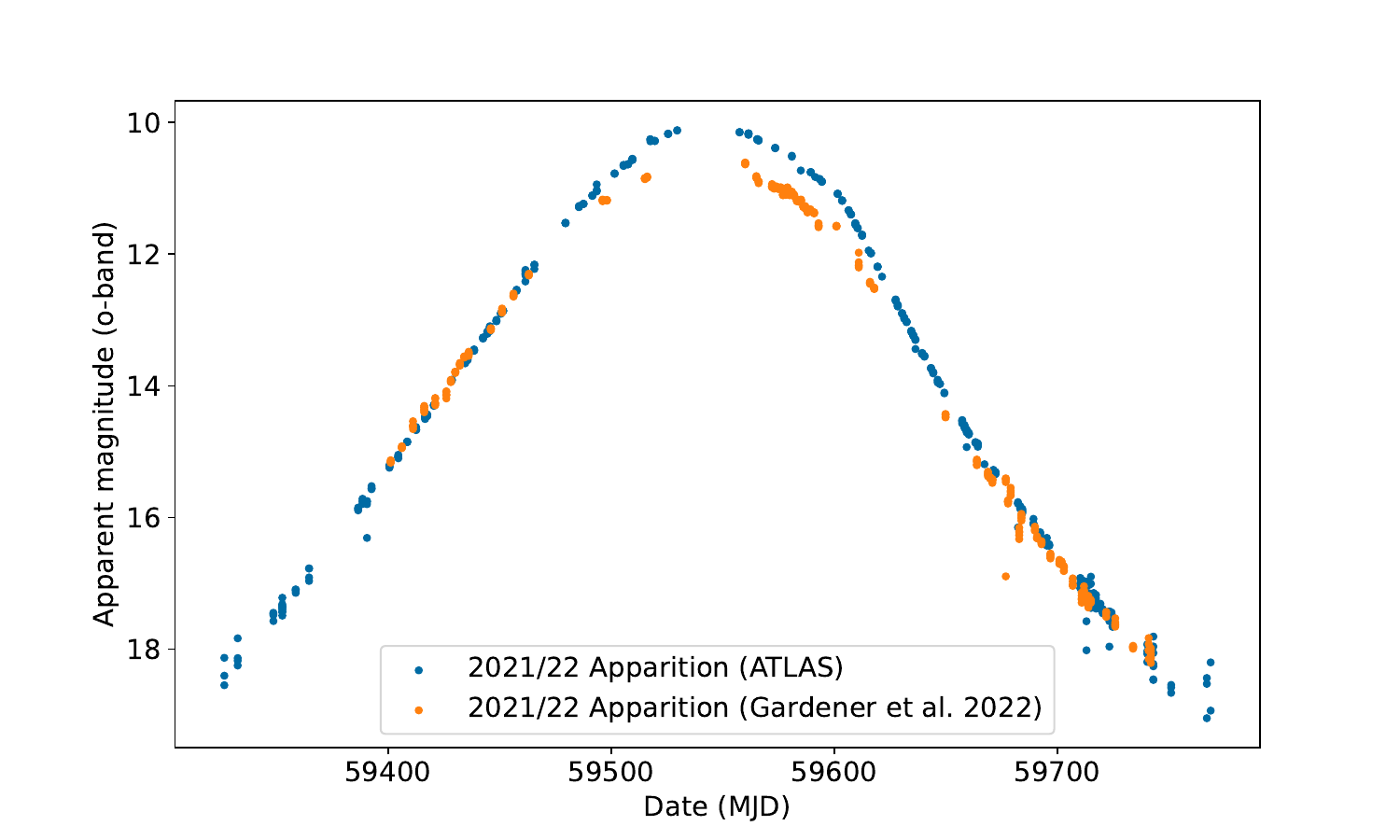}
    \caption{Comparison between our 67P o-band magnitudes transformed to $r(PS1)$ and and those obtained by \cite{2022MNRASGardener}. Good agreement is seen pre-perihelion until MJD $\sim 59500$ and post-perihelion after MJD$ \sim 59650$.
    }
    \label{fig:67P_2021mags}
\end{figure}

Given the large number of independent studies of 67P during the Rosetta mission, \cite{2022MNRASGardener} (hereafter G22) re-reduced and measured the above and other datasets using a bespoke photometry pipeline to provide self-consistent magnitudes for all available imaging data. They also reported $r(PS1)$ magnitudes for the 2021/2022 apparition that we studied. Hence, we first inter-compare those measurements with our own. Throughout this apparition, we measured a median color of $(c-o)=0.42$. From the photometric transformations given by \cite{2018PASPTonry}, this gives $r(PS1)\simeq o(ATLAS)+0.15$.  Both datasets are shown in Figure~\ref{fig:67P_2021mags}. There is a very good agreement at epochs before $MJD< 59500$ and after $MJD > 59650$. Between these dates, when the comet was brightest and near perihelion ($MJD=59520$), the G22 magnitudes are up to $\sim0.5$ magnitudes fainter than ours.

The reason for this discrepancy is not obvious. The G22 data were obtained with the 2.0 m Liverpool Telescope on the island of La Palma, which suffered a major volcanic eruption during this period that could affect photometry from the telescopes on the island. However, the eruption only lasted for 85 days during $MJD=59576-59561$ \citep{Carracedo2022},  matching the start of the discrepancy but ending $\sim90$ days too early. We also note that the pipeline employed by Gardener et al. used field stars in the science images for calibration, which should negate most affects of additional atmospheric extinction due to aerosol particles above the observatory. On the other hand, these dates coincide with the heliocentric distances  at which gas emission was most prominent in the previous apparition \citep{2017MNRASOpitom}. It is possible that extra gas emission in the o-band filter, perhaps due to the red CN (B--X) emission bands at $\lambda\simeq7700-8200$\AA \ , gave higher fluxes than the SDSS-\emph{r\textquotesingle} filter used by Gardener et al. However, the $(c-o)$ color of the comet did not change during this period, which is suprising if there was significant gas emission increasing the measured flux by $\sim50$\%. Finally, G22 noted that their magnitudes at this time were below those expected from modelling of previous apparitions. Inspection of their Figure~8 implies our data are close to their nominal predictions.

\begin{figure}[h!]
    \centering
    \epsscale{1.25}
    \plotone{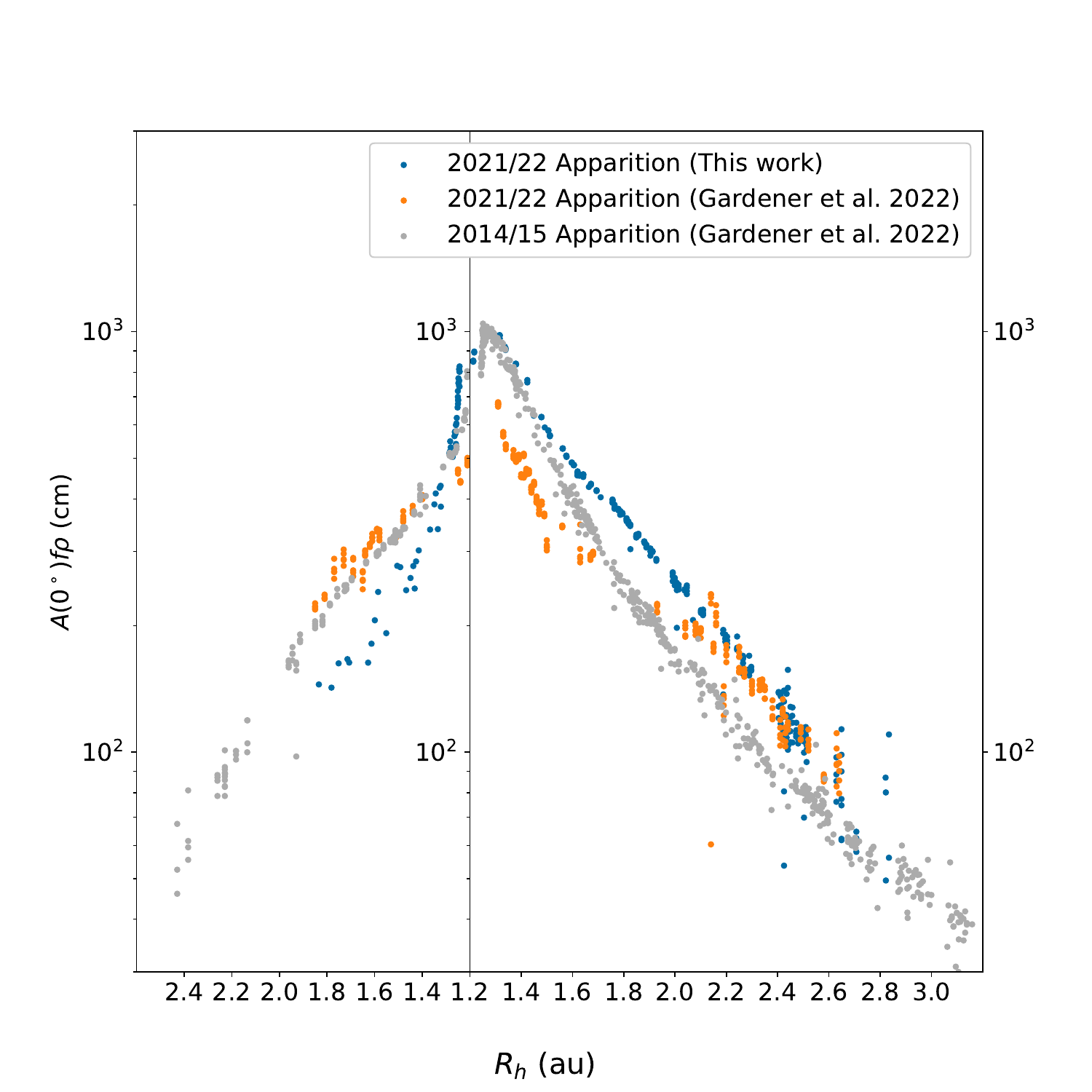}
    \caption{Comparison between our measurements of $A(0^\circ)f\rho$ for 67P in 2021-2022, values calculated from the magnitudes reported by \cite{2022MNRASGardener}, and values for the 2014/15 apparition calculated from the average $r_{PS1}$ lightcurve also reported by \cite{2022MNRASGardener}. As both those datasets reported magnitudes in the PS1 system, we have converted our $A(0^\circ)f\rho$ to that system as explained in the text. The apparent discontinuity at perihelion in the 2014-2015 values is due to the change in perihelion distance between the two apparitions.
    }
    \label{fig:67Pcomparison}
\end{figure}

To compare our 2021/2022 measurements with the average 2014/2015 $r(PS1)$ lightcurve in G22, we calculated $A(0^\circ)f\rho$ from their magnitudes. To do this, we converted our values for ATLAS \emph{o}-band into $r(PS1)$ values. Moving from an original photometric filter to a new (primed) filter, the change in $Af\rho$ will be given by:
\begin{equation}\label{eq:photo_systems_convert}
\frac{Afp'}{Afp}= 10^{0.4[(m_\odot'- m_\odot)+ (m_{com} - m_{com}')]}
\end{equation}
Here $m_\odot'$ and $m_\odot$ are the apparent magnitudes of the Sun in the original and new filter, taking care to choose the AB magnitude $r(PS1)=-26.93$ given by \cite{WillmerApjS2018} as appropriate. The cometary magnitudes in the original and new filter are $m_{com}$ and $m_{com}'$, where the difference will be given by the color-dependent transformation between the systems. The resulting evolution of $A(0^\circ)f\rho$ derived from the 2014/15 G22 magnitudes, together with the 2021-22 measurements from G22 and us, is shown in Figure~\ref{fig:67Pcomparison}.

This inter-comparison shows a number of different features. First, as expected from  Figure~\ref{fig:67P_2021mags}, our $A(0^\circ)f\rho$ values are higher than for the  G22 dataset around perihelion. Second, our absolute values within $\pm 0.2$ au of perihelion closely follow the derived activity in the 2014-15 apparition, including the rapid rise to peak activity followed by a slower fall-off. Third, at heliocentric distances $>1.4$ au pre-perihelion and  $>2.0$ au post-perihelion,  $A(0^\circ)f\rho$ in 2021-22 from both studies are consistently higher than in 2014-15.

Given that previous studies concluded the activity evolution was very similar from orbit to orbit \citep{Snodgrass2016}, the question is what could cause the enhanced activity seen at larger distances in Figure~\ref{fig:67Pcomparison}. One possibility is that the adopted phase function does not represent the scattering properties of dust in 67P. Looking at the phase angles during the two apparitions, 67P was at a higher phase angle when pre-perihelion in 2021 than 2015, while the situation was reversed post-perihelion. Therefore if this was the case, one might expect the discrepancy to change signs as well, with one curve being higher than the other pre-perihelion, and the reverse happening post-perihelion. This is not seen in Figure~\ref{fig:67Pcomparison}. Additionally, the dust phase function for 67P has been measured using {\it in-situ} Rosetta data in several studies. Early investigations close to the comet nucleus indicated a much steeper phase function than the Marcus-Schleicher 
function \citep{Bertini2017,Fulle2018,Fink2018}. However, it was subsequently found at distances $>100$ km from the nucleus that the phase function was similar to those measured remotely for other comets \citep{Bertini2019}. To confirm this we tried several phase functions measured by this latter study, but none gave a good agreement between the 2014/15 and 2021/22 apparitions.

Overall, while we agree with previous authors that the activity of 67P is broadly consistent between apparitions, it appears the behavior of the comet has changed at large distances. As noted by \cite{Boehnhardt2024}, 67P underwent a distant encounter with Jupiter of 0.31 au in 2018, changing the orientation of the perihelion vector by $5.3^\circ$ and reducing the perihelion distance by 0.03 au. It is unlikely that this perturbation induced significant changes to the spin-axis or surface structure. This change is more likely due to normal nucleus evolution and is not linked to its dynamical evolution.

\subsubsection{71P/Clark}
\citet{2009IcarFink} observed a decrease in dust activity from $A(0^\circ)f\rho = 363$ cm at $R_{h}$ = 1.55 au on May 29 1995, to $A(0^\circ)f\rho = 311$ cm at $R_{h}$ = 1.57 au Jun 24 1995. Due to solar conjunction, we lack measurements of 71P/Clark near perihelion where \citet{2009IcarFink} obtained them. We measured  $A(0^\circ)f\rho$ = 349 $\pm$ 4 cm at our smallest recorded heliocentric distance of $R_{h} = 1.81$ au on May 2 2023. Our measured $A(0^\circ)f\rho$ is comparable to that of the earlier study, but from our measurements, the activity is declining from a peak that likely occurred while the comet was obscured during solar conjunction. Although we lack measurements from the same heliocentric distances as those reported by \citet{2009IcarFink}, the higher measured $A(0^\circ)f\rho$ values in our dataset imply that 71P/Clark may be displaying enhanced dust activity during this orbit compared to its behaviour $\sim$ 5 orbits ago. 

\citet{2001A&ALowry} determined an $A(0^\circ)f\rho$ upper limit of $\leq$ 3.3 cm using the 4.2 m William Herschel Telescope in December 1998 at an $R_h = 4.40$ au inbound. We do not have measurements of 71P/Clark beyond $R_h$ = 3.1 au in this study, hence a comparison is not possible.

\citet{2009A&ALamy} used HST snapshot observations in the F675W filter to measure $A(0^\circ)f\rho = 92 \pm 1$ cm for 71P/Clark inbound at $R_h$ = 2.72 au in March 2000. We obtained measurements at $R_h = 2.74$ au of $A(0^\circ)f\rho = 16 \pm 5$ cm in the ATLAS \emph{c}-band pre-perihelion on 2022 April 28. There is clearly a large discrepancy between the two studies and the most significant factor contributing to this is likely the different apertures used to make the measurements. \citet{2009A&ALamy} used a circular aperture of 200 km radius compared to our radius of 10,000 km. It is also possible that the comet is not as active as early at this heliocentric distance on the most recent orbit. However, this comet did become reasonably active on this orbit, reaching a maximum $A(0^\circ)f\rho$ = 349 $\pm$ 4 cm when emerging from solar conjunction.

\subsubsection{80P/Peters-Hartley}
\citet{2009IcarFink} observed 80P/Peters-Hartley at \mbox{$R_{h} = 1.63$ au} and measured $A(0^\circ)f\rho = 185$ cm in June 1990. We measured $A(0^\circ)f\rho$ = 761 $\pm$ 5 cm at $R_{h} = 1.63$ au. The activity appears to have increased overall compared to its 1990 epoch. We also observed a very large increase in $A(0^\circ)f\rho$ in the measurements after solar conjunction. It is unclear if this JFC had undergone an outburst as we lack more data at this time and did not find any reports in the literature.

\subsubsection{81P/Wild 2}
Comet 81P/Wild 2 was the target of the \emph{Stardust} mission where the spacecraft collected thousands of cometary dust grains in 2004 and returned them to Earth \citep{2006SciBrownlee}. In June 1991, \citet{2009IcarFink} measured an average of $A(0^\circ)f\rho$ = 783 cm at $R_{h} = 2.25$ au. We measured $A(0^\circ)f\rho = 284$ $\pm$ 2 cm at $R_{h} = 2.3$ au on 2023 June 14. \citet{2009IcarFink} also measured the $Af\rho$ of 81P/Wild 2 on the next apparition, from $A(0^\circ)f\rho = 1726$ cm at $R_{h} = 1.85$ au on 1997 January 28 to $A(0^\circ)f\rho = 922$ cm at $R_{h} = 1.61$ au on 1997 June 3. We have measured  $A(0^\circ)f\rho = 871$ $\pm$ 6 cm post-perihelion at $R_{h} = 1.62$ au. The two measurements are not consistent which is again may be due to the different methods used to measure $Af\rho$ between these studies.

\subsubsection{96P/Machholz 1}
96P/Machholz 1 is unique among short-period comets with an orbit of extremely high eccentricity, $e$ = 0.96, low perihelion distance, $q$ = 0.12 au, and high inclination, $i$ = 57.5\textdegree. 96P/Machholz 1 also has a unique chemical composition with the lowest CN-to-OH ratio of any known comet but a high $C_{2}$-to-OH ratio \citep{Langland-Shula2007, 2008AJSchleicher}. The $C_{2}$-to-CN ratio of 96P/Machholz 1 was the highest known value of any comet when measured by \citet{2008AJSchleicher}. The origin of 96P/Machholz 1 still remains uncertain, where its origin may have been either a region in our solar system from which the probability is very low for crossing inside of Neptune’s orbit, or it is of interstellar origin \citep{Langland-Shula2007, 2008AJSchleicher}.

\citet{2008AJSchleicher} measured $Af\rho$ in the UV, blue and green continuum. We compared our results with the green continuum $Af\rho$ measurements as they are made at the closest wavelength to the central wavelength of the \emph{o}-band in ATLAS. \citet{2008AJSchleicher} measured a range of $A(0^\circ)f\rho$ from $A(0^\circ)f\rho$ = 2.1--2.8 cm from May to June 2007 at $R_{h}$ = 1.3--1.7 au respectively. We have not made any measurements of 96P/Machholz 1 below $R_{h}$ = 1.47 au, however our measurements at $R_{h}$ = 1.7 au give $A(0^\circ)f\rho$ $\sim$ 75 cm where the comet is active.

\citet{2019AJEisner} shows that 96P/Machholz 1 can appear inactive at heliocentric distances as low as $R_{h}$ = 2.3 au, suggesting strong seasonal effects or volatile depletion in the upper layers of the nucleus due to the repeated perihelion passages at an extremely low heliocentric distance. In our images from ATLAS, the comet does not clearly show a coma until $R_{h}$ $<$ 1.9 au, and our $A(0^\circ)f\rho$ measurements appear to be nucleus dominated at distances larger than this, in agreement with that work.

\subsubsection{103P/Hartley 2}
Comet 103P/Hartley 2 was the target of the extended \emph{Deep Impact} mission \emph{EPOXI} (Extrasolar Planet Observation and Deep Impact Extended Investigation) which performed a close flyby in 2010 \citep{2011SciAHearn}. \citet{2009IcarFink} measured an increase in $Af\rho$ of $A(0^\circ)f\rho = 172$ cm at $R_{h} = 1.26$ au on 1997 October 31 to $A(0^\circ)f\rho = 319$ cm near perihelion at $R_{h} = 1.17$ au on 1998 January 31. We measured $A(0^\circ)f\rho = 73.6 \pm 0.4$ cm on 2023 August 23 at $R_{h} = 1.26$ au in the \emph{c}-band, decreasing to $A(0^\circ)f\rho = 64.4 \pm 0.3$ cm at $R_{h} = 1.17$ au on 2023 September 7 in the \emph{o}-band. However, our measurements imply significant gas emission in the ATLAS \emph{c}-band at this time and therefore should be interpreted with caution. 

103P/Hartley 2 appeared active in measurements by \cite{2003A&ALowry} with an $A(0^\circ)f\rho = 61 \pm 4$ cm at $R_{h} = 4.57$ au outbound in June 1999. We do not have measurements beyond $R_{h}$ = 2.5 au for 103P/Hartley 2 so a comparison between the two studies is not possible.\citet{2011A&ALara} measured  $A(0^\circ)f\rho$ $\sim$ 320 cm within a circular aperture of radius $\rho$ $\sim$ 4700 km in the red continuum (RC) filter at $R_{h}$ = 1.06 au using the William Herschel Telescope.
 In the ATLAS \emph{o}-filter, which has the closest central wavelength, we measured an $A(0^\circ)f\rho = 113.7 \pm 0.4$ cm. The two studies do not agree, possibly caused by the difference in apertures.

\subsubsection{104P/Kowal 2}
\citet{2009IcarFink} measured a decrease in dust production from $A(0^\circ)f\rho$ = 134 cm at $R_{h} = 1.72$ au on 1997 November 29 to $A(0^\circ)f\rho = 59$ cm at $R_{h} = 1.42$ au on 1999 March 25. We measured an increase in dust production from $A(0^\circ)f\rho = 6 \pm 1$ cm at $R_{h}$ = 1.73 au on 2021 September 24 to $A(0^\circ)f\rho = 13 \pm 1$ cm in the ATLAS \emph{c}-band. Both of our measurements are lower than that of the comparison study and again, may be due to the \citet{2009IcarFink} measurements being obtained spectroscopically, however the difference between the two studies is substantial.
\cite{2003A&ALowry} measured a 3$\sigma$ upper limit of $A(0^\circ)f\rho$ $\leq$ 7.4 cm at $R_{h}= 3.94$ au outbound on its orbit. We do not have measurement beyond $R_{h} = 2.15$ au and therefore comparison is not possible.

\subsubsection{113P/Spitaler}
113P/Spitaler was observed by \citet{2023PSJGicquel} at $R_{h}$ = 2.34 au with an $A(0^\circ)f\rho = 42\pm 11$ cm in December 2014 in the NEOWISE data. We observed 113P/Spitaler from 2021 August 18 to 2023 March 2nd in the ATLAS images. We measured an $A(0^\circ)f\rho$ = 55 $\pm$ 13 cm at $R_{h}$ = 2.36 au in the ATLAS data. The $A(0^\circ)f\rho$ values we have measured for 113P/Spitaler in the ATLAS images differ significantly from night to night, however the measurements are consistent within 1$\sigma$ between the two studies. This is somewhat surprising, given the NEOWISE data were obtained in the W1 band at $\lambda=$ 3.4 $\mu m$ and with a slightly smaller aperture radius of 8200 km. However this seeming agreement may be due to the relatively large uncertainties on these measurements.

\subsubsection{117P/Helin-Roman-Alu 1}
We observed 117P/Helin-Roman-Alu 1 from 2019 December 24 to 2024 February 2 in the ATLAS images. Comet 117P/Helin-Roman-Alu 1 was observed by NEOWISE on two separate visits as discussed in \citet{2023PSJGicquel}, who measured $A(0^\circ)f\rho$ = 757 $\pm$ 170 cm at $R_{h}$ = 3.06 au in May 2014. We observed this comet at $R_{h}$ = 3.06 au with a measured $A(0^\circ)f\rho$ = 1240 $\pm$ 10 cm. Our measurement of $A(0^\circ)f\rho$ at the same heliocentric distance as the comparison study agrees within 3$\sigma$. \citet{2023PSJGicquel} also observed the comet in October 2014, finding $A(0^\circ)f\rho$ = 1008 $\pm$ 226 cm  at $R_{h}$ = 3.22 au. We do not have observations of 117P/Helin-Roman-Alu 1 at $R_{h}$ = 3.22 au, however, we have observations at 3.1 au and 3.4 au where $A(0^\circ)f\rho$ = 847 $\pm$ 21 cm and 537 $\pm$ 32 cm respectively. Our measurements at 3.1 au agree also within 1$\sigma$.

\subsubsection{287P/Christensen}
We observed 287P/Christensen from 2022 June 10 to 2024 Jan 7th in the ATLAS images. It was measured with NEOWISE by \citet{2023PSJGicquel} at $R_{h}$ = 3.08 au with an $A(0^\circ)f\rho$ = 141 $\pm$ 35 cm in October 2014 in the NEOWISE data. We measured an $A(0^\circ)f\rho$ = 33 $\pm$ 5 cm at $R_{h}$ = 3.08 au in the ATLAS data, hence our measurements do not agree. This may imply that the comet was much less active on this apparition. Indeed, over our entire set of observations, the maximum measured $A(0^\circ)f\rho$ was $\sim$ 57 cm which is significantly lower than the comparison study.\\

\subsubsection{Summary of the comparative analysis with previous studies}

Most of the $A(0^\circ)f\rho$ measurements in this study were consistent with previous studies within 3$\sigma$. However, several comets show notable discrepancies. These differences were largely attributed to a combination of factors. First, measurements made at different heliocentric distances naturally reflect differences in the measured activity between the two studies.  Second, secular changes in JFC activity patterns between epochs may also have resulted in differing $A(0^\circ)f\rho$ measurements for the same comet. For example, we compared $A(0^\circ)f\rho$ measurements of 19P/Borrelly over thirty years (\mbox{$\sim$4 orbits}) later. Additionally, differences in the bandpasses used between our study and comparison studies might lead to small offsets in measured $A(0^\circ)f\rho$ values due to the wavelength dependent albedo of the dust grains. These factors collectively underscore the importance of considering observational context when interpreting cometary activity. However, the overall agreement within 3$\sigma$ with our measured dust production rates imply no large changes in $A(0^\circ)f\rho$ between orbits for the majority of JFCs.\\

\subsection{Outbursts} 
Cometary outbursts are sudden releases of mass from the cometary nucleus which can be triggered by several different mechanisms, causing the comet to typically increase in brightness by 2--3 magnitudes \citep{1990QJRASHughes}. For example, fractures on the surface on comets can drive cliff collapses as seen on 67P/Churyumov-Gerasimenko which are usually triggered by the associated or nearby activity \citep{2017SciEl-Maarry}. Another potential cause of an outburst may be crystallisation of amorphous ices phase transition which release trapped subsurface gases \citep{2017MNRASAgarwal, 2022arXivPrialnik} or the nucleus splitting and fragmenting. 

In Paper I we identified 6 outbursts in 5 comets.
From our ATLAS observations of the 2022 and 2023 perihelion JFCs, we measured an additional six outbursts in five different comets. The outbursts were identified by visual inspection of the JFC light curves and activity curves. 
Table~\ref{tab:outbursts} lists the duration of the outbursts in days, the date the outburst occurred in our ATLAS data, the $R_h$ distance or range the outburst spanned, and the change in magnitude from the quiescent level of activity. Four of the six outbursts detected had been previously reported: 97P/Metcalf-Brewington \citep{2021ATelKelley97P}, 99P/Kowal 1 \citep{2021ATelKelley99P}, 285P/LINEAR \citep{2022ATelKelley285P} and 382P/Larson \citep{2021ATelKelley382P}. 

From our four years of observing JFCs in the ATLAS data, 
we identified  12 outbursts in 10 comets, meaning $\sim 9$\% of JFCs exhibit clear outbursts per perihelion passage. The average increase in brightness was $1.3 \pm 0.8$ magnitudes. The outbursts ranged from an increase in magnitude from $-0.2$ to $-2.8$ which agrees with those seen on other comets \citep{1955ApJWhitney, 2008A&ATrigo, 2022IcarWesolowski}. However, these results are influenced by observational biases, as the smallest detected outburst we measured had an increase of $-0.2$ magnitudes. If an outburst produces only a small brightness increase and the SNR for the JFC is low, the outburst may be indistinguishable from normal fluctuations in brightness or fall within the range of photometric uncertainties, making detection more challenging or impossible. The primary observational bias in the ATLAS data is the sensitivity to small outbursts with brightness increases of $<$0.2 mag. Similar short-lived outbursts have also been observed in other JFCs, such as those recorded during the Deep Impact mission \citep{2005SciAHearn}.

\begin{table*}
\centering
\caption{JFC outburst date range and the increase in magnitude from the 2022 and 2023 JFC subset. $\Delta$Magnitude is measured from before the outburst to the maximum magnitude achieved. I = inbound (pre-perihelion), O = outbound (post-perihelion). Comet 118P/Shoemaker-Levy 4 experienced two outbursts at different epochs represented by (1) and (2).}
\begin{tabular}{lcccc}
\hline
Comet & Outburst dates start & Outburst duration  & $R_h$ & $\Delta$Magnitude\\
 & & (days) & (au) &\\
\hline
97P/Metcalf-Brewington & 2021 Oct 29  &  14 & 2.7$^{I}$ & $-2.2$\\
99P/Kowal 1 & 2021 May 12  &  15 & 4.9$^{I}$ & $-0.5$\\
118P/Shoemaker-Levy 4 (1)& 2022 Sep 25 & 25 & 1.9$^{I}$ & $-0.2$\\
118P/Shoemaker-Levy 4 (2)& 2022 Nov 27 & 23 & 1.8$^{O}$ & $-0.5$\\
285P/LINEAR & 2022 Aug 1 & 46 & 2.1 - 2.4$^{I}$ & $-2.1$\\ 
382P/Larson & 2021 Sep 23 & 36 & 4.5$^{I}$ & $-1.4$\\ 
\hline
\end{tabular}
\label{tab:outbursts}
\end{table*}

\section{Conclusions} \label{sec:conclusion}
We analyzed high-cadence observations of 74 active JFCs reaching perihelion in 2022 and 2023 using wide-field ATLAS survey data. We  measured $A(0^\circ)f\rho$ for a total of 116 JFCs from 2020--2023 following our previous work in Paper I. Over the four years of ATLAS observations, this yielded a total of 34,046 images for detailed analysis. The results are as follows: \\

\begin{enumerate}[nolistsep]
\item We measured the relative dust production rates of the JFCs using the $Af\rho$ parameter and its variation as a function of heliocentric distance. When corrected to zero phase angle, the activity in this sample spans $\sim$ 2 orders of magnitude in $A(0^\circ)f\rho$ , showing the wide range in dust production among comets from the JFC population.

\item We found a number of JFCs reaching their maximum $A(0^\circ)f\rho$ pre-perihelion, likely due to seasonal effects. Despite this, there remained a clear preference for JFCs to reach their maximum $A(0^\circ)f\rho$ post-perihelion, which we attributed to a combination of slow-moving larger dust grains and the observational viewing geometry. Notably, several comets reached their maximum $A(0^\circ)f\rho$ more than 100 days after perihelion; 99P/Kowal 1, 170P/Christensen, 244P/Scotti, 254P/McNaught, 408P/Novichonok-Gerke and P/2020 WJ5 (Lemmon). These comets are predominantly characterized by lower eccentricities and therefore likely much more gradual heating of the nucleus and ices. The time of perihelion relative to maximum activity may therefore be of lower significance for these comets. 

\item We found that all comets in our sample — including those that reached their maximum $A(0^\circ)f\rho$ more than a year after perihelion — exhibited this maximum within 10\% of their orbital period. This agrees with expectations that all JFCs reach their maximum dust production close to perihelion.

\item The majority of the JFCs in our sample reach their maximum $A(0^\circ)f\rho$ within 2.5 au of the Sun, where water-ice sublimation likely dominates their activity. However, several comets exhibit significant activity at larger distances (408P/Novichonok-Gerke, 254P/McNaught, 244P/Scotti, P/2019 LD2 (ATLAS), 9P/Kowal 1 and P/2020 WJ5 (Lemmon)). Their activity may be driven by more volatile species such as CO, CO$_{2}$, or NH$_{3}$ \citep{2017PASPWomack}, but our data cannot reveal the nature of the activity driver.

\item The activity index, $n$ was calculated for each JFC. The pre-perihelion activity indices ranged from $-12.5$ to 5.5, and $-8.0$ to 1.3 post-perihelion. Most of the JFCs in our sample possessed negative activity indices, but there were a few that possess positive values. This was attributed to the challenge of accurately measuring activity where the variation in $A(0^\circ)f\rho$ is smaller than the measurement uncertainties. Overall, the measured activity indices pre-perihelion exhibit a much broader range than those post-perihelion, with greater variation among the number of JFCs sharing a similar activity index. The average activity index for the four-year sample pre-perihelion was $-5.2 \pm 4.5$ and post-perihelion was $-3.2 \pm 2.7$. Including the JFCs from all four years of our study and increasing the sample size from 20 to 56 JFCs with measurable $n$, we find that the distribution of activity indices is statistically different between pre-perihelion and post-perihelion. 

\item To determine the true intrinsic range of $A(0^\circ)f\rho$ between JFCs, we interpolated and extrapolated $A(0^\circ)f\rho$ to a common distance of 2 au. We found that there was no observable correlation between the intrinsic activity of the JFCs and their perihelion distance.

\item For 444P/WISE-PANSTARRS and 459P/Catalina in the 2022-23 sample, there was a comet detection but no visible coma in any of the images for a range of dates. We measured upper limits for the nuclear radii which were $R_{n} \leq 1.5\pm0.2$ km, and $R_{n} \leq 1.7\pm0.1$ km respectively. 

\item We compared our measurements of 67P/Churyumov-Gerasimenko in 2021-22 with the extensive data published by \cite{2022MNRASGardener}, both for that apparition and 2014-15. A similar pattern of activity was seen for all three datasets, but there were quantitative differences in the calculated $A(0^\circ)f\rho$. At face value the data imply the comet exhibited higher activity in parts of the orbit than previously measured in earlier apparitions.

\item In our analysis of 116 JFCs, we identified  outbursts in 10 comets, an occurrence rate of $\sim 9$\% per perihelion passage. As we are limited by sensitivity to outbursts brighter than $-0.2$ magnitudes, this is a lower limit to the rate of observable outbursts in JFCs. The outbursts increase in magnitude ranged from $-0.2$ to $-2.8$, with an average increase of $-1.3\pm0.8$ magnitudes. 
 
\end{enumerate}

The ATLAS survey provides high-cadence observations for significant portions of a comet's orbit in the inner solar system, which is ideal for observing JFCs when they are most active. However, combining ATLAS with deeper surveys such as Pan-STARRS \citep{2016Chambers} or LSST \citep{2023ApJSSchwamb} would allow us to observe JFCs over a much greater fraction of their orbits, potentially to aphelion. Although these deeper surveys can reach fainter objects, they do not offer the same high cadence in a single filter as ATLAS and therefore they might be used in conjunction with ATLAS to provide a more comprehensive orbital coverage. 
By integrating optical coverage from several surveys, a more comprehensive picture of the dust activity throughout JFC orbits can be achieved.

\begin{acknowledgments}
We thank the two anonymous reviewers for useful comments and feedback, both of which have improved this paper. This work has made use of data from the Asteroid Terrestrial-impact Last Alert System (ATLAS) project. ATLAS is primarily funded to search for near earth asteroids through NASA grants NN12AR55G, 80NSSC18K0284, and 80NSSC18K1575; byproducts of the NEO search include images and catalogs from the survey area. The ATLAS science products have been made possible through the contributions of the University of Hawaii Institute for Astronomy, the Queen's University Belfast, the Space Telescope Science Institute, the South African Astronomical Observatory (SAAO), and the Millennium Institute of Astrophysics (MAS), Chile. A.F.G. acknowledges support from the Department for the Economy (DfE) Northern Ireland postgraduate studentship scheme. A.F. acknowledges support from STFC award ST/T00021X/1.
\end{acknowledgments}
\emph{Facility:} ATLAS (Chile, Haleakala, Mauna Loa and South Africa telescopes) 

\emph{Software:} Astropy \citep{{Astropy2013}, {price2018astropy}, {2022ApJAstropyCollab}}, Jupyter Notebook \citep{Kluyver2016jupyter}, Matplotlib \citep{Hunter:2007}, Numpy \citep{harris2020array}, Pandas \citep{mckinney2010data}, Photutils \citep{larry_bradley_2022_6825092}, Python (https://www.python.org), SciPy \citep{2020SciPy-NMeth}.

\section*{Data Availability}
The photometric data used in this project is available on Zenodo under an open-source Creative Commons Attribution license: 


\bibliography{main}{}
\bibliographystyle{aasjournal}

\figsetstart \label{JFC_figset}
\figsetnum{1}

\figsetgrpend
\figsetend

\end{document}